\documentclass[useAMS,usenatbib,usegraphicx]{mn2e}

\usepackage{times}

\newcommand{\iraf}  {{\sc iraf}}
\newcommand{\BaI} {Ba\,{\sc i}}
\newcommand{\BaII} {Ba\,{\sc ii}}
\newcommand{\CaI} {Ca\,{\sc i}}

\newcommand{\CrI} {Cr\,{\sc i}}

\newcommand{\kms}{km\,s$^{-1}$}
\newcommand{\masyr}{mas\,yr$^{-1}$}

\newcommand{\Ergsi}{erg\,s$^{-1}$cm$^{-2}$}
\newcommand{\vsini}{$V_{rot}\sin i$}

\newcommand{\CIII} {C\,{\sc iii}}

\newcommand{\FeI} {Fe\,{\sc i}}

\newcommand{\HeI}  {He\,{\sc i}}
\newcommand{\HeII} {He\,{\sc ii}}

\newcommand{\MgI} {Mg\,{\sc i}}

\newcommand{\NIII} {N\,{\sc iii}}

\newcommand{\Halpha} {H$\alpha$}
\newcommand{\Hbeta}  {H$\beta$}
\newcommand{\Hgamma} {H$\gamma$}

\newcommand{\lam}{$\lambda$}

\title[Spectroscopy of the cataclysmic variable BF Eri]
{BF Eridani: a cataclysmic variable with a massive white dwarf and an evolved secondary}

\author[V.\,V.\,Neustroev and S.\,Zharikov]{V.\,V.\,Neustroev$^{1}$\thanks{E-mail:
vitaly.neustroev@nuigalway.ie} and S.\,Zharikov$^2$\\
$^{1}$Centre for Astronomy, National University of Ireland, Galway, Newcastle Rd., Galway, Ireland\\
$^{2}$Instituto de Astronom{\'\i}a, Universidad Nacional Aut\'onoma de M\'exico, Apartado Postal 877,
22830, Ensenada, Baja California, M\'exico\\
}
\begin{document}

\date{Accepted 2008 January 4.  Received 2008 January 1; in original form 2007 December 3}

\pagerange{\pageref{firstpage}--\pageref{lastpage}} \pubyear{2007}

\maketitle

\label{firstpage}

\begin{abstract}
  We present high- and medium-resolution spectroscopic observations of the cataclysmic variable BF~Eri during its low
  and bright states. The orbital period of this system was found to be 0.270881(3) days. The secondary star is clearly
  visible in the spectra through absorption lines of the neutral metals \MgI, \FeI\ and \CaI. Its spectral type was
  found to be K3$\pm$0.5. A radial velocity study of the secondary yielded a semi-amplitude of $K_2=182.5\pm0.9$ \kms.
  The radial velocity semiamplitude of the white dwarf was found to be $K_1=74\pm3$ \kms\ from the motion of the
  wings of the \Halpha\ and \Hbeta\ emission lines. From these parameters we have obtained that the secondary in BF~Eri
  is an evolved star with a mass of 0.50--0.59 $M_\odot$, whose size is about 30 per cent larger than a
  zero-age main-sequence single-star of the
  same mass. We also show that BF~Eri contains a massive white dwarf ($M_1\geq1.23 M_\odot$), allowing us to consider
  the system as a SN Ia progenitor. BF~Eri also shows a high $\gamma$-velocity
  ($\gamma\ =-94$ \kms) and substantial proper motion. With our estimation of the distance to the system
  ($d\approx700\pm200$ pc), this corresponds to a space velocity of $\sim$350 \kms\ with respect to the dynamical local
  standard of rest. The cumulative effect of repeated nova eruptions with asymmetric envelope ejection might explain the
  high space velocity of the system. We analyze the outburst behaviour of BF~Eri and question the current classification
  of the system as a dwarf nova. We propose that BF~Eri might be an old nova exhibiting ``stunted'' outbursts.
\end{abstract}


\section{Introduction}

  Cataclysmic Variables (CVs) are close interacting binaries that contain a white dwarf (WD) accreting material from
  a companion, usually a late main-sequence star. CVs are very active photometrically, exhibiting variability on time
  scales from seconds to centuries (see review by \citealt{Warner}). Dwarf novae (DNs) are an important subset of CVs.
  Observable features that distinguish DN from other CVs such as nova-like variables (NLs) are the recurrent outbursts
  of 2--6 mag that occur on timescales of days to years. In this paper we concentrate on BF Eridani, a little-studied
  dwarf nova candidate.

  BF Eri was discovered and first classified as a slowly varying variable star by \citet{HanleyShapley}. During
  the following half a century no observations of the star were reported. As a consequence, the specific nature of
  the star remained unknown. The identification of the Einstein survey source 1ES 0437-046 with BF~Eri \citep{Elvis}
  and optical spectroscopy by \citet{Schachter} eventually led to its correct identification as a CV. Later, the ROSAT
  identification of the Einstein source confirmed the classification of BF~Eri as a CV of an unknown type \citep{Chisholm}.

\begin{table*}
\label{ObsTab}
\begin{center}
%\begin{flushleft}
\caption{Log of observations of BF~Eri}
\begin{tabular}{ccclcccll}
\hline\hline
Set     &  Date        & HJD Start  & Instrument &$\Delta\lambda$$^a$& $\lambda$~range & Exp.Time & Number   & Duration\\
        &              &  2450000+  &            &  (\AA)            & (\AA)           &  (sec)   & of exps. &  (hours)\\
\hline
2005-N1 & 2005-Oct-30  &  3673.982  & B\&Ch$^b$& 6.0          & 3680--6770      &  300     & 16       & 1.44   \\
2005-N2 & 2005-Oct-31  &  3674.821  & B\&Ch   &  6.0          & 6025--8000      &  300     & 40       & 3.50   \\
2005-N3 & 2005-Nov-01  &  3675.821  & B\&Ch   &  2.1          & 6145--7225      &  300     & 58       & 5.25   \\
\hline
2006-Nov \\
2006-N1 & 2006-Nov-22  &  4061.884  & Echelle & 0.234         & 3915--7105      & 1200     & 9        & 3.13   \\
2006-N2 & 2006-Nov-23  &  4062.728  & Echelle & 0.234         & 3915--7105      & 1200     & 17       & 6.81   \\
2006-N3 & 2006-Nov-24  &  4063.710  & Echelle & 0.234         & 3915--7105      & 1200     & 19       & 7.08   \\
2006-N4 & 2006-Nov-25  &  4064.684  & Echelle & 0.234         & 3915--7105      & 1200     & 3        & 0.70   \\
        &              &  4064.879  & Echelle & 0.234         & 3915--7105      & 1200     & 3        & 0.70   \\
2006-N5 & 2006-Nov-26  &  4065.698  & Echelle & 0.234         & 3915--7105      & 1200     & 20       & 7.26   \\
2006-N6 & 2006-Nov-27  &  4066.689  & Echelle & 0.234         & 3915--7105      & 1200     & 14       & 4.78   \\
\hline
2006-Dec \\
        & 2006-Dec-13  &  4082.840  & Echelle & 0.234         & 3915--7105      & 1200     & 2        & 0.35   \\
        & 2006-Dec-14  &  4083.850  & Echelle & 0.234         & 3915--7105      & 1200     & 2        & 0.35   \\
        & 2006-Dec-15  &  4084.808  & Echelle & 0.234         & 3915--7105      & 1200     & 2        & 0.36   \\
\hline
\end{tabular}
%\end{flushleft}
%\end{center}
\begin{tabular}{l}
%\begin{flushleft}
$^a$ -- $\Delta\lambda$ is the FWHM spectral resolution \\
$^b$ -- B\&Ch - Boller \& Chivens spectrograph  \\
\end{tabular}
\end{center}
%\end{flushleft}
\end{table*}

  The long term light curve was first analysed by \citet{Watanabe} who detected two outbursts and concluded that
  BF~Eri is a DN with a recurrence period of 40--50 days and the observed magnitude range between 13.2 and fainter
  than 14.7 mag. \citet{Kato99} and \citet{KatoUemura} confirmed this classification but mentioned the low amplitude
  of its outbursts and the difference in the outburst activity between seasons as the amplitudes of outbursts were
  remarkably different -- 1.0 mag and 1.6 mag. \citet{Kato99} was unable to detect any orbital variability.

  Prior to the beginning of this work there was no other study of the system in the literature\footnote{
%  When this work was almost ready, a new paper on BF~Eri has appeared.
  \citet{BF-Eri} have recently presented optical medium-resolution spectroscopy of BF~Eri, obtained primarily to
  determine the orbital period. Below we discuss their results.}.
  This motivated us to perform time-resolved
  spectroscopy of BF~Eri in order to study its properties in more detail. In this paper we present and discuss the first
  high- and medium-resolution spectroscopic observations of BF~Eri during its low and bright states in 2005 and 2006.
  In Section~\ref{ObsSec} we  describe our  observations and  data reduction. The data  analysis and the results
  are presented  in Section~\ref{DatAnSec}. Here we find the orbital period, measure the radial velocities of the WD
  and the secondary star, and determine the spectral type of the secondary and its rotational velocity.
  In Section~\ref{DopMapSec}
  we discuss the results of Doppler tomography of the emission lines from BF~Eri. The system parameters are obtained
  in Section~\ref{SysParSec}, while a discussion and a summary are given in Sections~\ref{DiscSec} and \ref{SumSec},
  respectively.

%************************  Average Spectra  ************************************
\begin{figure*}
\centering
\includegraphics[width=17cm]{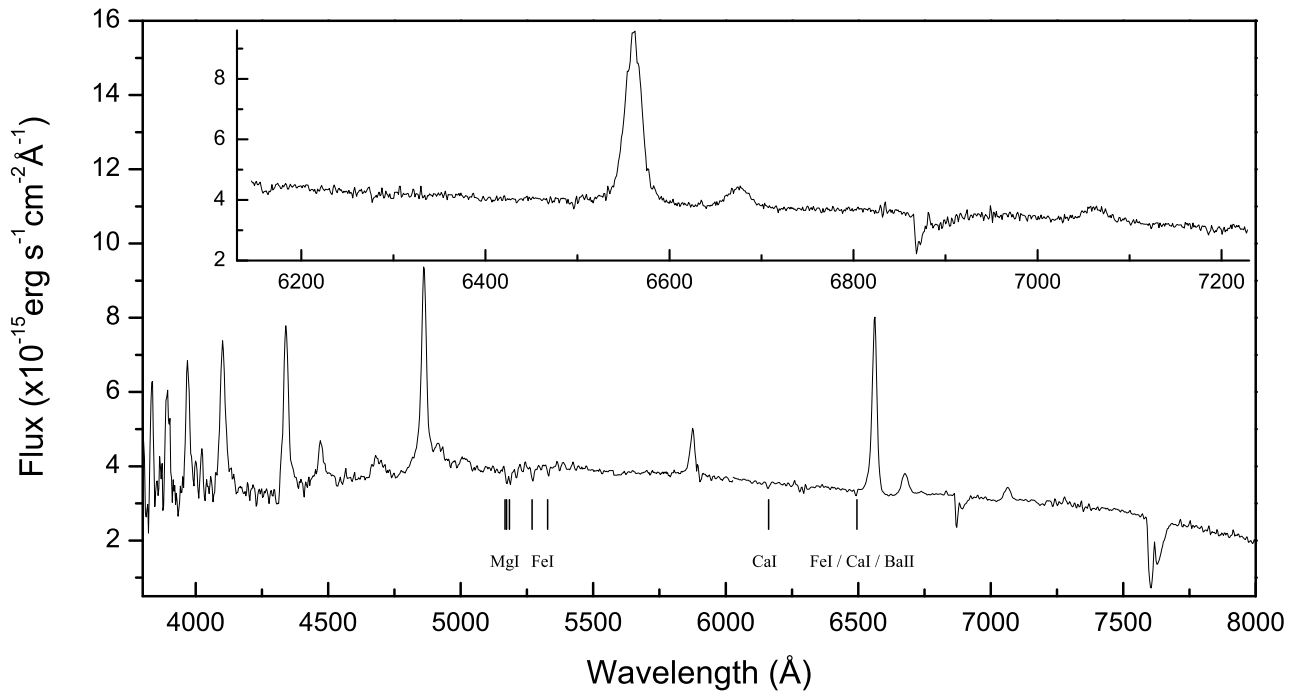}
\caption{The average spectra of BF~Eri from the \textit{2005} set, uncorrected for orbital motion. The lower
spectrum is combined from the \textit{2005-N1} and \textit{2005-N2} sets while the inset spectrum is
an average of all spectra from the \textit{2005-N3} set.}
\label{aver_spec}
\end{figure*}
%************************  Average Spectra  ************************************

\section{Observations and Data reduction}
\label{ObsSec}

  The spectra presented here were obtained at the Observatorio Astronomico Nacional
  (OAN SPM) in Mexico on the 2.1~m telescope. The first observations, in 2005, were conducted with the
  Boller \& Chivens spectrograph, equipped with a 24~$\mu$m ($1024\times1024$) SITe CCD.
  Observations were made during three consecutive nights of October 30--31 -- November 1 in the wavelength
  range of 3680--6770 {\AA}, 6025--8000 {\AA} and 6145--7225 {\AA} respectively.
  Spectra were obtained in the first order of 400 line/mm (for the first two nights) and 1200 line/mm
  (for the third night) gratings with corresponding spectral resolutions of around 6.0 {\AA} and 2.1 {\AA}
  respectively. A total of 114 spectra were obtained with 300~s individual exposures, half of them
  on the third night. Cu-He-Ne-Ar lamp exposures were taken before and after the observations of the target for
  wavelength calibrations, and the standard spectrophotometric stars Feige110, HR3454 and G191-B2B \citep{Oke}
  were observed for flux calibrations.

  In order to check the current photometric state of the object, during the
  first night we obtained a few photometric observations on an accompanying 1.5~m telescope at the same site.
  The V magnitude was about 15.0--15.2, indicating that BF Eri was definitely in the quiescent state.
  These observations showed that BF~Eri is a long-period system and that the obtained data were
  not sufficient to determine the orbital period with good precision.

  Further observations were obtained during 6 consecutive nights of November 22--27, 2006
  and 3 nights of December 13--15, 2006 with the Echelle spectrograph \citep{Echelle} attached
  to the same 2.1~m telescope. This instrument gives a resolution of
  0.234 \AA\ pixel$^{-1}$ at H$\alpha$ using the UCL camera and a CCD-Tek chip of
  1024$\times$1024 pixels with a 24~$\mu$m$^2$ pixels size. The spectra cover 25 orders
  and span the spectral range 3915--7105 \AA. A total of 85 spectra were obtained in November
  and 6 spectra in December with 1200~s individual exposures.
  In November we also took spectra of the spectral type templates Gliese 4.1 (G5V),
  WO 9808 (G6V), NN4366 (G7V), Gliese 793.1 (G9V), Gliese 75 (K0V), Gliese 68 (K1V), Gliese 33 (K2V), Gliese 183 (K3V),
  Gliese 53.1 (K4V), Gliese 69 (K5V), Gliese 169 (K7V), V~Psc (M1V) and Gliese 844 (M2V).
  All the nights of observations were photometric and the seeing ranged from 1 to 2 arcsec.
  Table \ref{ObsTab} provides a journal of the BF~Eri observations.

  The reduction procedure was performed using \iraf. Comparison spectra of Th-Ar
  lamps were used for the wavelength calibration.
  The absolute flux calibration of the spectra was achieved by taking nightly
  echellograms of the standard stars HD93521 and HR153.
  Though we used a wide slit (2\arcsec) with seeing usually noticeably less than the slit
  width, this does not warrant excellent flux calibration, since only an average curve for
  atmospheric extinction and a permanent E-W orientation of the slit were used.
  At the same time, due to an unexpectedly observed outburst of BF~Eri in the November set we
  found it useful to obtain some photometric and colour information from our spectra.
  For this we followed an approach which we used for investigating an outburst of
  the dwarf nova BZ~UMa \citep{Neustroev}. Namely, we defined an internal photometric
  system comprising four colour bands $u'\,b'\,v'\,r'$ centered at 4000\,\AA, 4550\,\AA,
  5590\,\AA, 6380\,{\AA} with widths of 50\,\AA, 100\,\AA, 120\,\AA, 150\,{\AA}
  respectively. The colour indices were calculated as $C=-2.5\log(f_1/f_2)$,
  where $f_1$ and $f_2$ are the fluxes averaged across the corresponding bands.
  To check the stability and the flux calibration accuracy we have determined
  the $u'-b'$, $b'-v'$ and $v'-r'$ colours of the control star HR153 using
  our nightly spectra, and compared these with its published spectral energy
  distribution. The colours did not differ by more than 0.03 mag between any of our
  observations, and our colours were within 0.06 mag of the published spectra.

  To improve the confidence of the results presented in this paper we also acquired the
  \textit{night-} and \textit{set}-averaged spectra obtained by means of co-adding of all spectra
  collected during each night and each set of observations respectively. Additionally we obtained
  the phase-averaged spectra for the \textit{2005-N3} and \textit{2006-Nov} datasets. For this we have phased the
  individual spectra with the orbital period, derived in Section~\ref{PeriodSec}, and then co-added the spectra
  into 15 and 17 separate phase bins respectively. In fact, only 11 bins were filled for the \textit{2005-N3} set
  because these data were taken without complete orbital coverage.

%************************  Average Spectra  ************************************
\begin{figure*}
\centering
\includegraphics[width=8cm]{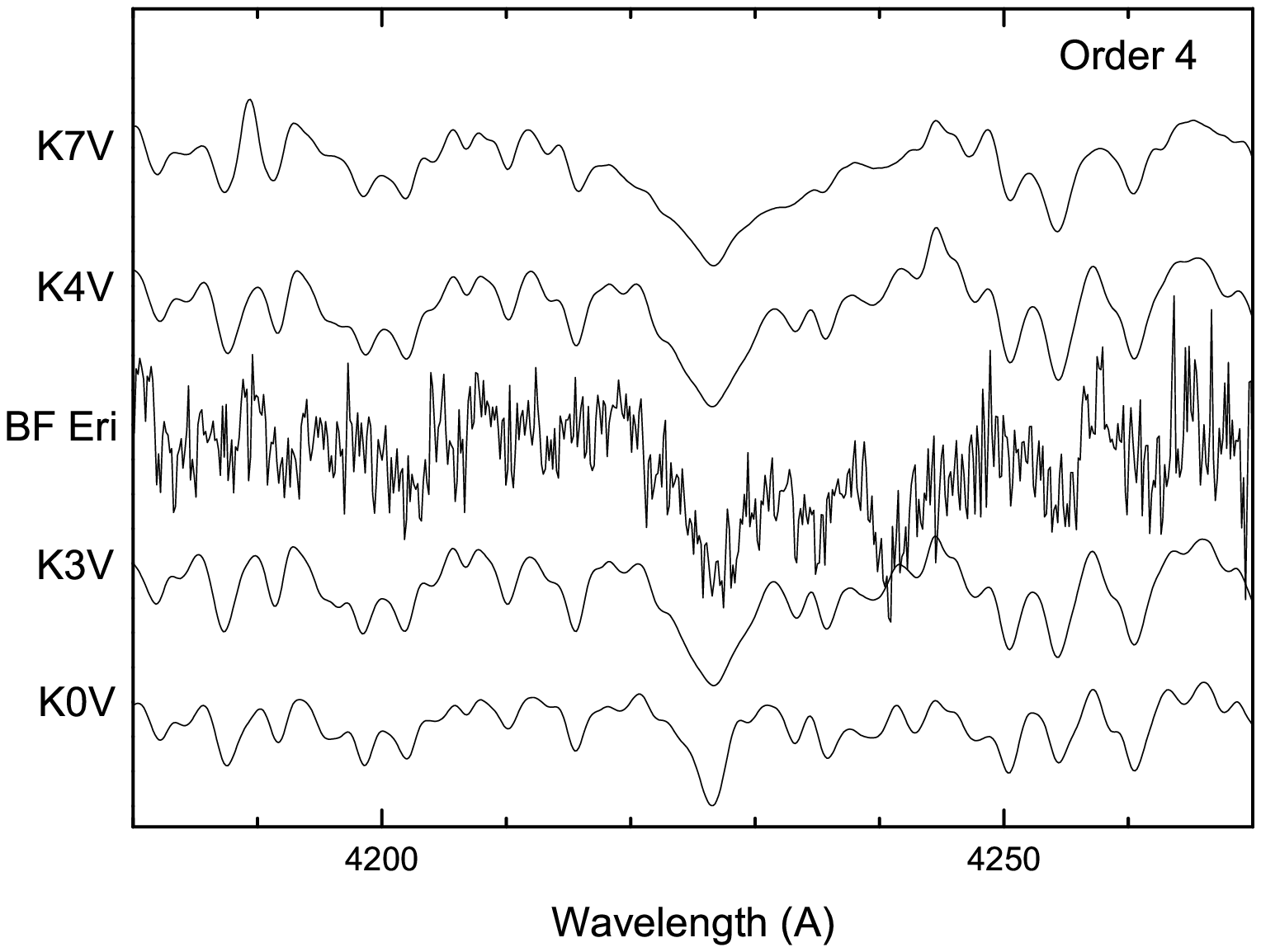} \includegraphics[width=8cm]{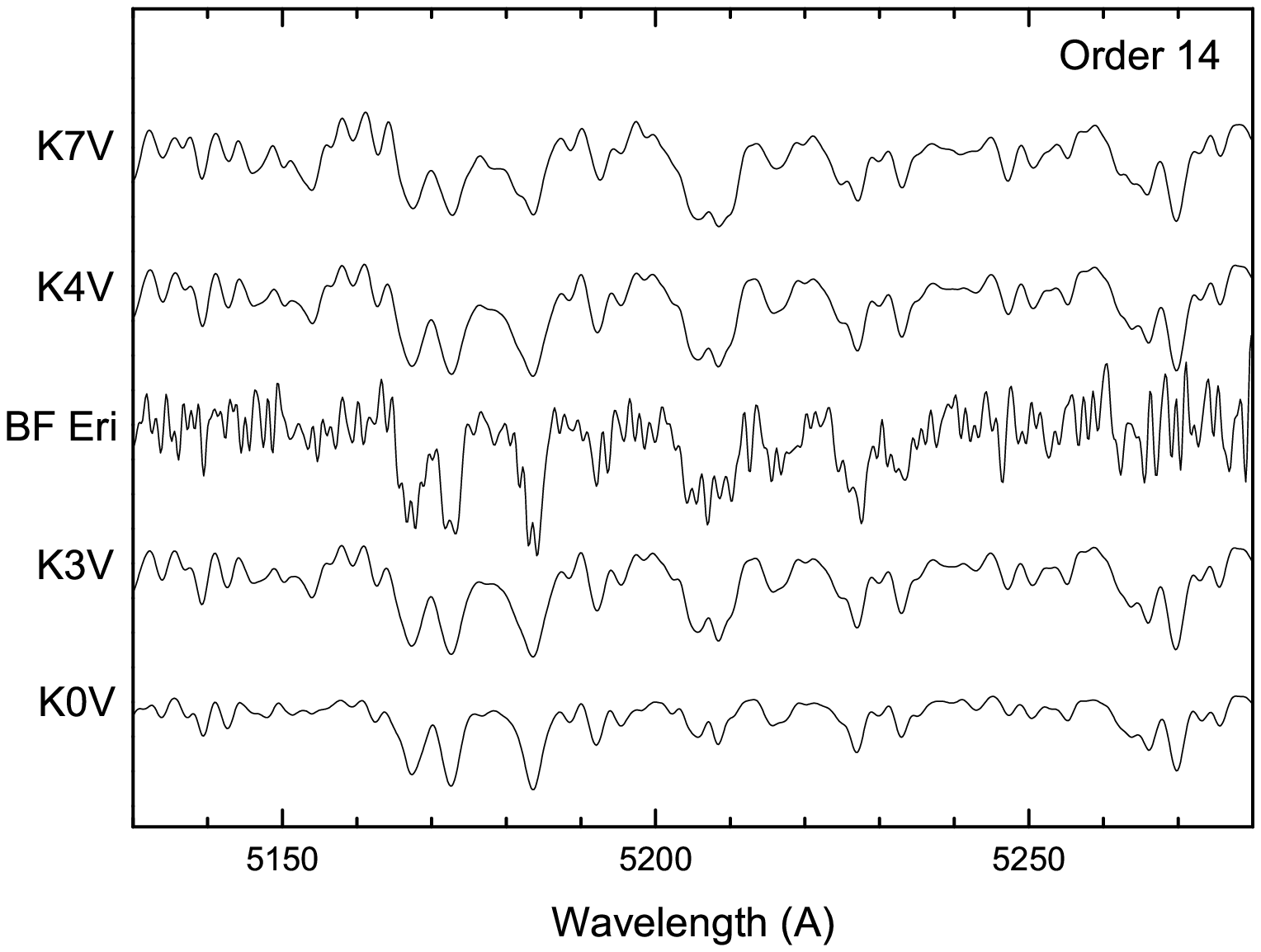} \\
\includegraphics[width=8cm]{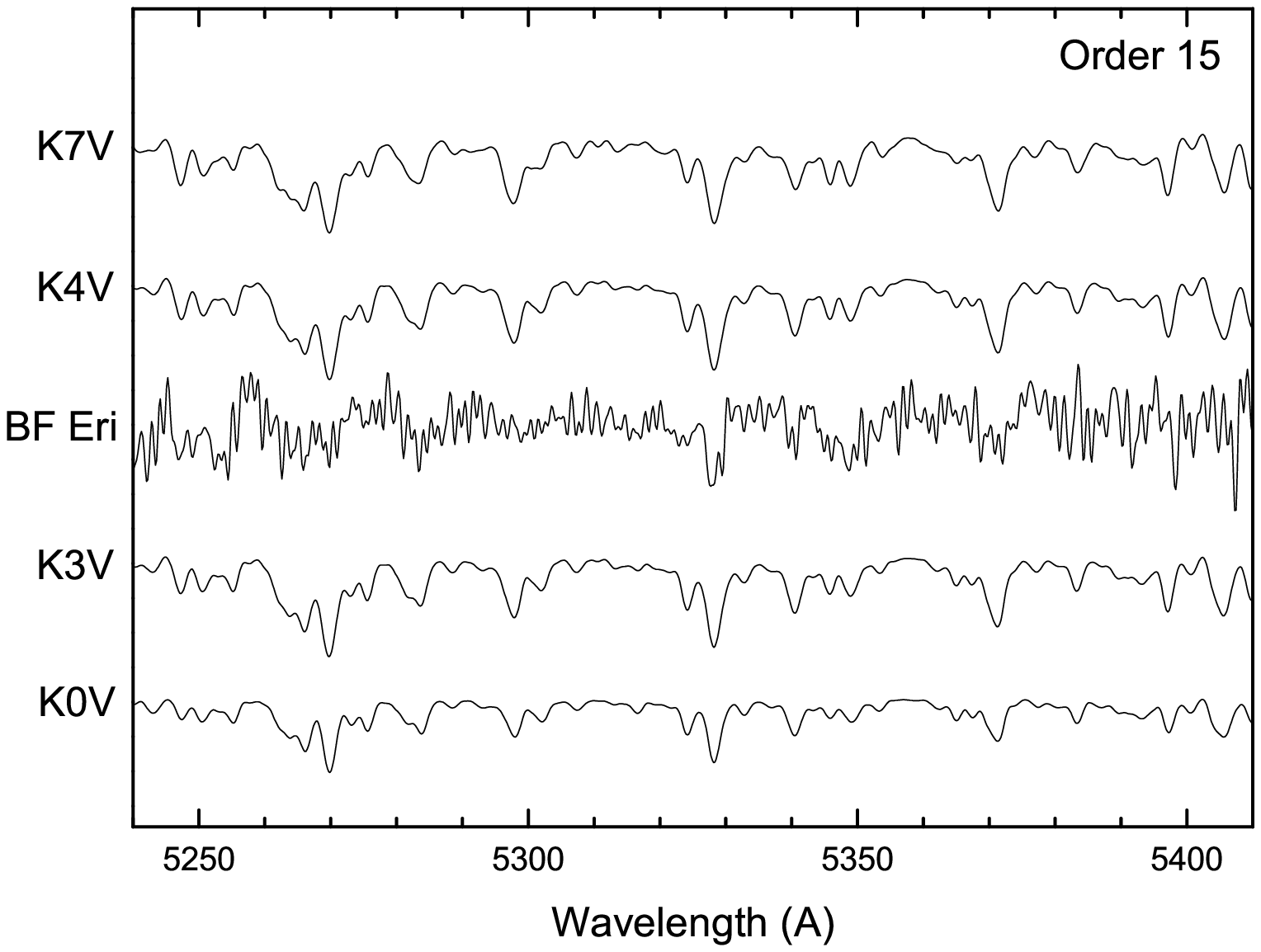} \includegraphics[width=8cm]{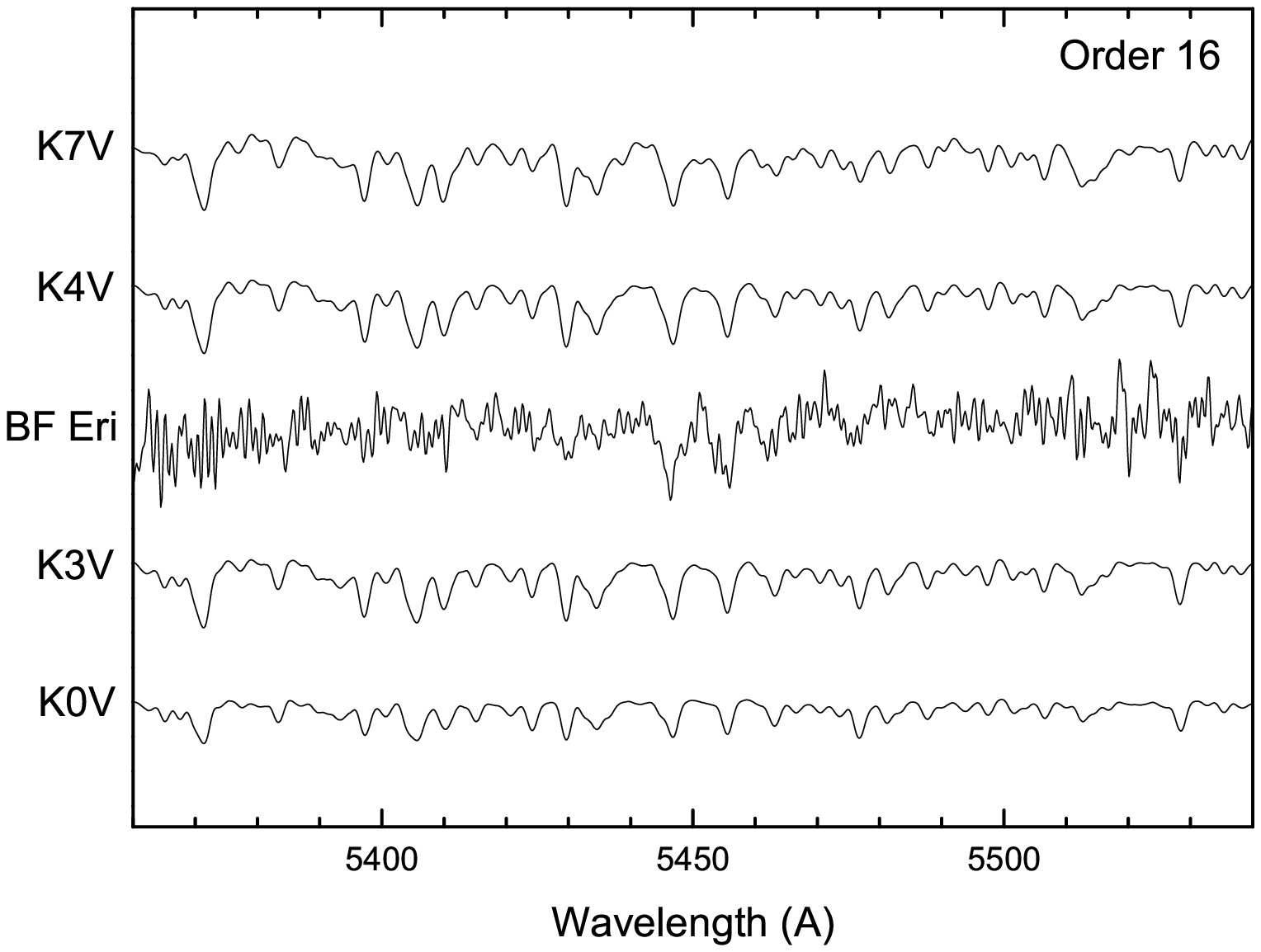} \\
\includegraphics[width=8cm]{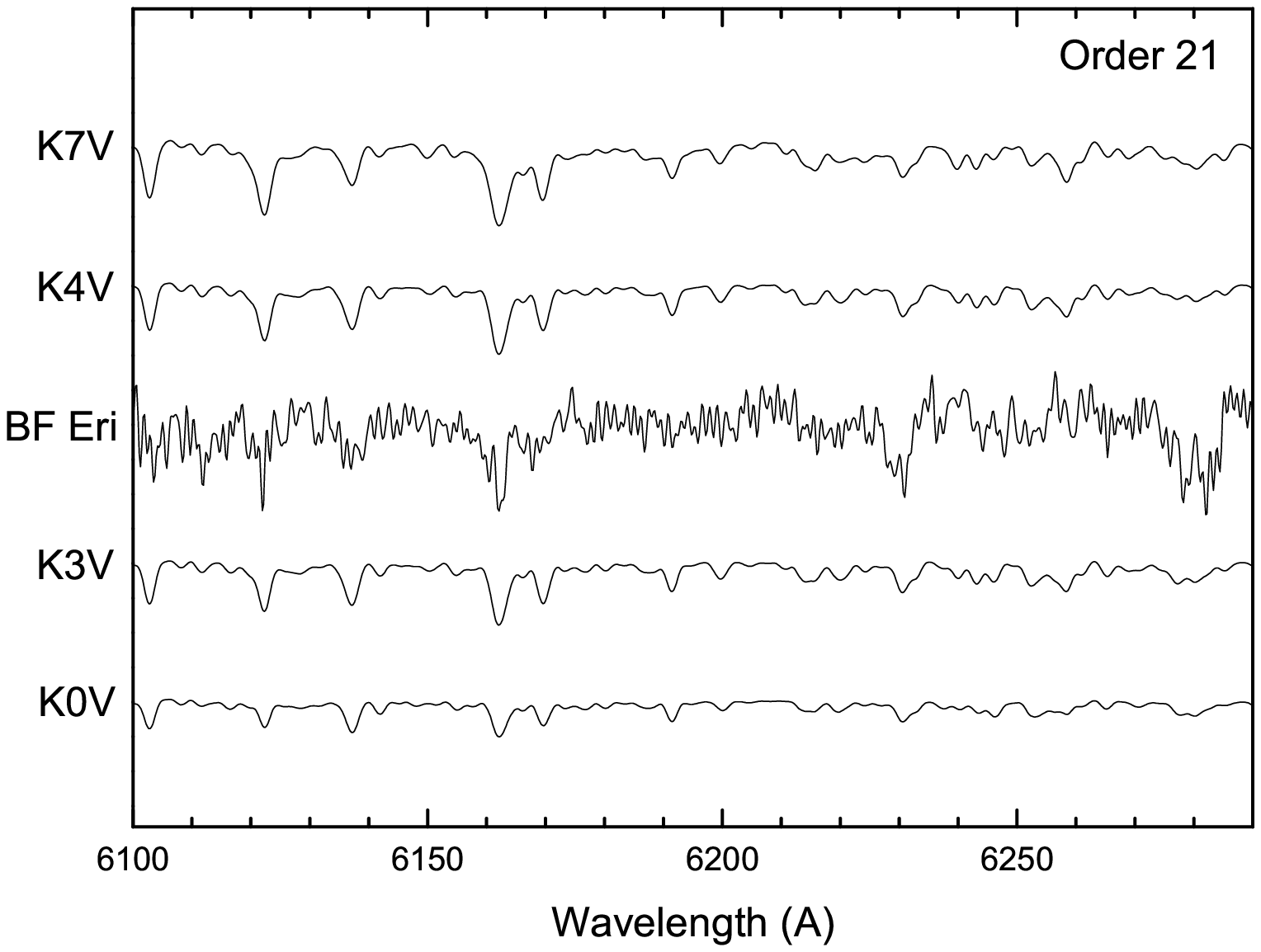} \includegraphics[width=8cm]{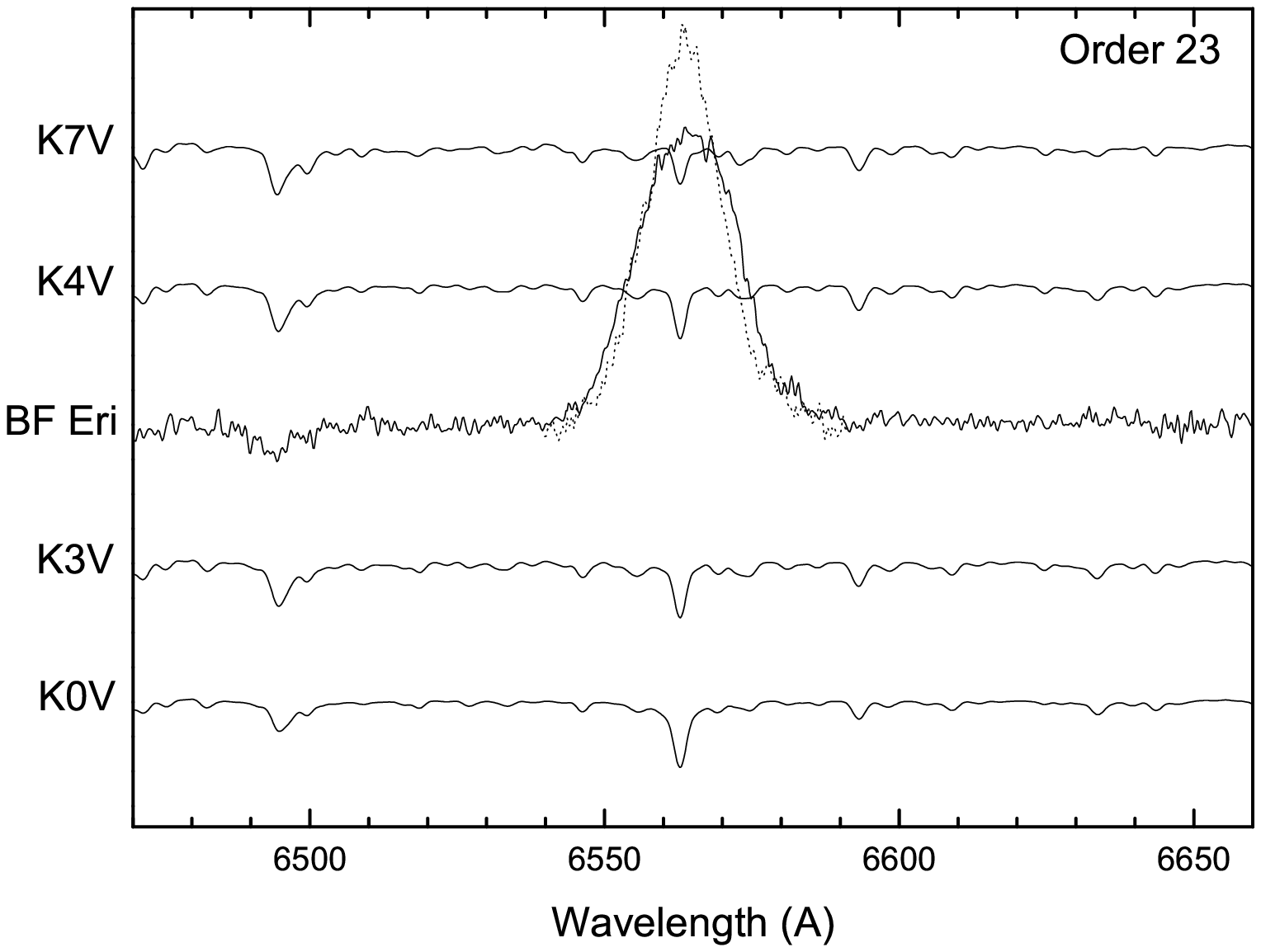} \\
\caption{The selected averaged and normalized spectral orders from the \textit{2006-Nov} set of Echelle
observations of BF~Eri, compared with several spectral standards. The spectra of BF~Eri have been corrected
for orbital motion of the secondary star. We also additionally show spectral order 23 which consists of the region
around the strongest emission line \Halpha. While the signal-to-noise ratios of the absoption lines were increased after
the Doppler correction, the emission line has been smeared out. To correct this we also show, by the dashed line,
the profile of \Halpha\ corrected for orbital motion of the WD. The spectra of standards have been broadened
by 78 \kms\ to account for the rotational broadening of the lines in BF~Eri and by 59 \kms\ to account for orbital
smearing of the BF~Eri spectra during exposure (see Section~\ref{RotVelSec} and footnote~\ref{rotvelfootnote}
 for details).}
\label{echelle_spec}
\end{figure*}
%************************  Average Spectra  ************************************

\section{Data analysis and results}
\label{DatAnSec}

\subsection{Average spectra}
\label{AveSpecSec}

    The mean spectra of BF~Eri are shown in Fig.~\ref{aver_spec}. These spectra are an average
    and a combination of all spectra from the \textit{2005} set, uncorrected for orbital motion.
    During these observations the system appears to have been in a quiescent state (see Section~\ref{LongTerm}).
    The spectrum is typical for a dwarf nova. It is dominated by strong and broad emission lines of
    hydrogen and neutral helium. In addition to them \HeII\ $\lambda4686$ is observed also but the
    \CIII/\NIII\ blend is not seen. All emission lines are single-peaked. The Balmer decrement is flat, indicating
    that the emission is optically thick as is normal in dwarf novae.
%    The secondary star is clearly visible even in these, uncorrected for orbital motion average spectra
%    as absorption lines of the neutral metals \MgI, \FeI\ and \CaI.
    The secondary star is clearly visible as absorption lines of the neutral metals \MgI, \FeI\ and \CaI,
    even in the average spectra, uncorrected for orbital motion.

    Much more detail can be seen in the Echelle spectra. Fig.~\ref{echelle_spec} shows the selected averaged and
    normalised spectral orders from the \textit{2006-Nov} set of observations. We have selected to show those spectral
    orders which exhibit the strongest absorption-line features used for the following analysis. These plots were
    Doppler-corrected into the frame of the secondary star, employing the ephemerides from Table~\ref{Tab:Syspar}.
    The absorption spectrum is rich in features. In particular we note the \MgI\ triplet at $\lambda\lambda$
    5167-5173-5184 \AA\ (order 14), the \FeI\ $\lambda$5328 line (order 15), the \CaI\ $\lambda$6162 line (order 21),
    the \FeI/\CaI/\BaII\ $\lambda$6495 blend (order 23).
    Spectral order 4, although very noisy, indicates the presence of the \CaI\ $\lambda$4226 line.
    We also additionally show order 23 which consists of the region around
    the strongest emission line \Halpha. While the signal-to-noise ratios of the absoption
    lines were increased after the Doppler correction, the emission line has been smeared out. To correct this,
    we also show, by the dashed line, the profile of \Halpha\ corrected for orbital motion of the WD.
    However note that even in this higher spectral resolution the emission line profile still appears to be
    single-peaked.

    Table ~\ref{LineParamTab} outlines fluxes, equivalent widths, velocity widths and relative intensities
    of the major emission lines measured from the averaged spectra of all the observational sets.

\begin{table*}
\begin{center}
\caption{Fluxes, Equivalent Widths (EW), Full Widths at Half Maximum (FWHM),
Full Widths at Zero Intensity (FWZI) and Relative Intensities of the major emission lines}
\begin{tabular}{llclccc}
\hline\hline
Set     & Spectral &  Flux            &   EW    & Relative  & FWHM          & FWZI          \\
        & line     &  ($\times10^{-14}$ \Ergsi) & (\AA)  & Intensity & (km s$^{-1}$) & (km s$^{-1}$) \\
\hline
2005-N1 & \Halpha  &  10.5            &  32.3   & 2.18      & 1000          & 4200          \\
        & \Hbeta   &  16.0            &  41.6   & 2.41      & 1320          & 6600          \\
        & \Hgamma  &  12.6            &  39.6   & 2.40      & 1500          & 4000          \\
        & \HeI\ $\lambda5875$ &  2.8  &   7.5   & 1.33      & 1100          & 4000          \\
        & \HeI\ $\lambda6678$ &  1.6  &   5.1   & 1.16      & 1430          & 3200          \\
        & \HeII\ $\lambda4686$&  2.8  &   7.7   & 1.14      &               &               \\
\hline
2005-N2 & \Halpha  &  11.3            &  33.3   & 2.46      &  915          & 3800          \\
        & \HeI\ $\lambda6678$ &  1.7  &   5.3   & 1.18      & 1300          & 3200          \\
        & \HeI\ $\lambda7065$ &  1.2  &   3.9   & 1.12      & 1270          & 3100          \\
\hline
2005-N3 & \Halpha  &  12.7            &  32.5   & 2.50      &  915          & 3850          \\
        & \HeI\ $\lambda6678$ &  1.8  &   4.9   & 1.18      & 1250          & 2800          \\
        & \HeI\ $\lambda7065$ &  1.3  &   4.0   & 1.13      & 1170          & 3100          \\
\hline
2006-N1 & \Halpha  &  17.6            &  22.5   & 2.30      &  710          & 2300          \\
        & \Hbeta   &  17.4            &  12.2   & 1.89      &  800          & 2350          \\
        & \HeI\ $\lambda5875$ &  3.6  &   3.6   & 1.22      &  870          &               \\
\hline
2006-N2 & \Halpha  &   9.8            &  15.7   & 1.97      &  670          & 1900          \\
        & \Hbeta   &   7.4            &   6.5   & 1.54      &  710          & 1650          \\
        & \HeI\ $\lambda5875$ &  1.7  &   2.0   & 1.17      &  670          &               \\
\hline
2006-N3 & \Halpha  &  11.0            &  18.2   & 1.99      &  670          & 2900          \\
        & \Hbeta   &   7.1            &   5.7   & 1.45      &  650          & 2300          \\
        & \HeI\ $\lambda5875$ &  2.0  &   2.2   & 1.18      &  740          &               \\
\hline
2006-N4 & \Halpha  &   8.9            &  15.5   & 1.92      &  660          & 2350          \\
        & \Hbeta   &   7.1            &   6.9   & 1.49      &  800          & 1800          \\
        & \HeI\ $\lambda5875$ &  2.0  &   2.6   & 1.16      &  690          &               \\
\hline
2006-N5 & \Halpha  &   9.3            &  16.3   & 1.92      &  650          & 2500          \\
        & \Hbeta   &   9.9            &  10.7   & 1.66      & 1050          & 2500          \\
        & \HeI\ $\lambda5875$ &  2.5  &   3.5   & 1.22      &  890          &               \\
\hline
2006-N6 & \Halpha  &  12.3            &  23.0   & 2.12      &  760          & 3000          \\
        & \Hbeta   &   8.2            &   8.5   & 1.55      &  960          & 2350          \\
        & \HeI\ $\lambda5875$ &  2.3  &   3.2   & 1.18      &  820          &               \\
\hline
2006-Nov& \Halpha  &  10.8            &  17.0   & 2.02      &  670          & 2250          \\
        & \Hbeta   &   8.3            &   7.5   & 1.51      &  830          & 2450          \\
        & \Hgamma  &   7.3            &   5.3   & 1.42      &  920          &               \\
        & \HeI\ $\lambda5875$ &  2.2  &   2.6   & 1.17      &  790          & 1600          \\
        & \HeII\ $\lambda4686$&  1.6  &   1.2   & 1.10      &               &               \\
\hline
2006-Dec& \Halpha  &  17.4            &  26.0   & 2.46      &  730          & 2550          \\
        & \Hbeta   &  19.3            &  25.0   & 2.35      &  950          & 2700          \\
        & \HeI\ $\lambda5875$ &  5.6  &   8.2   & 1.35      & 1430          & 2000          \\
\hline
\end{tabular}
\label{LineParamTab}
\end{center}
\end{table*}

\subsection{Long term photometric and spectral changes}
\label{LongTerm}

\begin{figure}
    \centering
     \includegraphics[width=8.0cm]{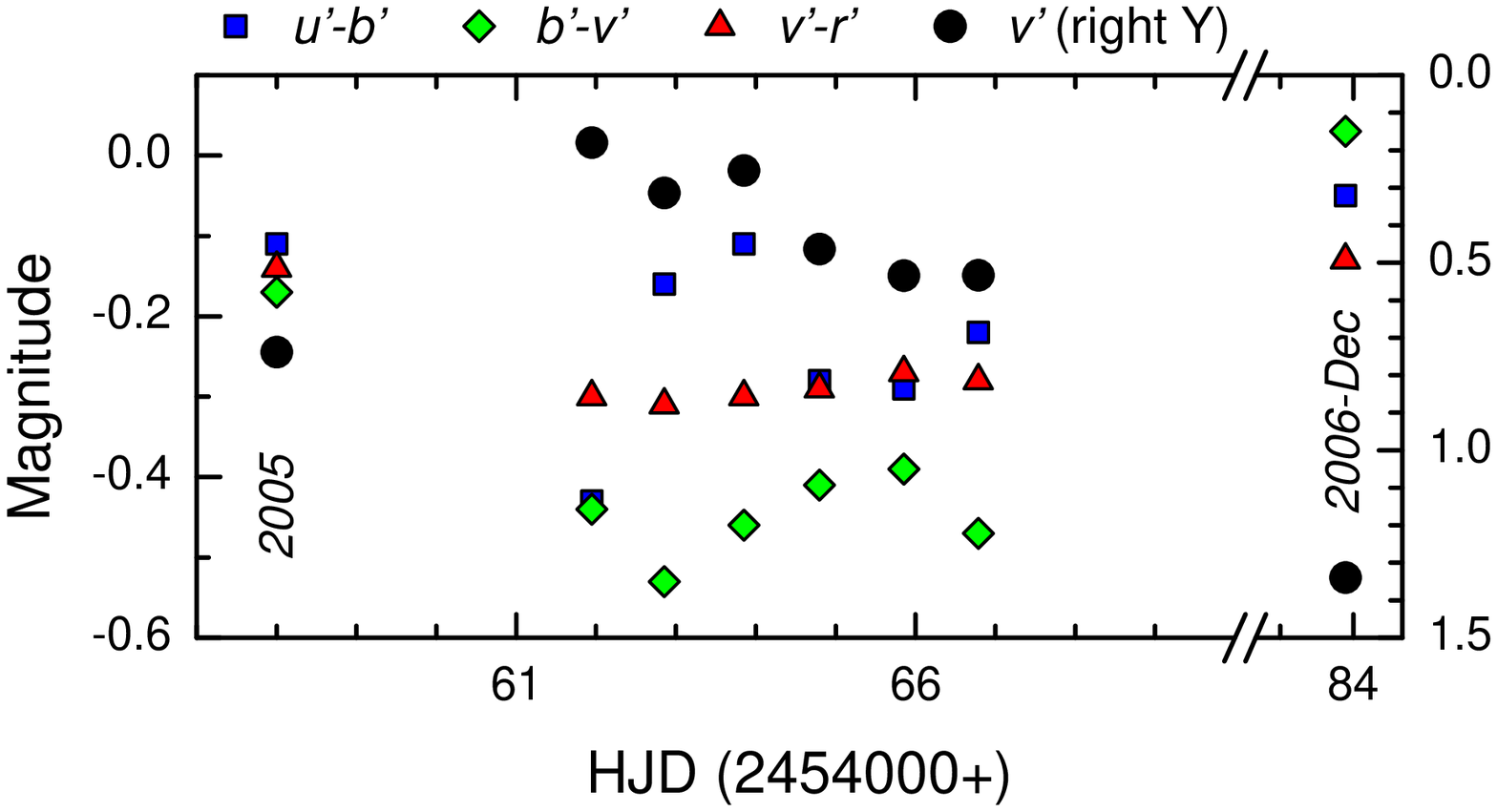}
       \caption{The $u'-b'$, $b'-v'$ and $v'-r'$ colour indices observed in BF~Eri.
        Symbols represent data obtained from the \textit{night-}averaged spectra.
        The $v'$ light curve is also shown (dots, right Y).}
          \label{colours}

    \centering
    \includegraphics[width=8.0cm]{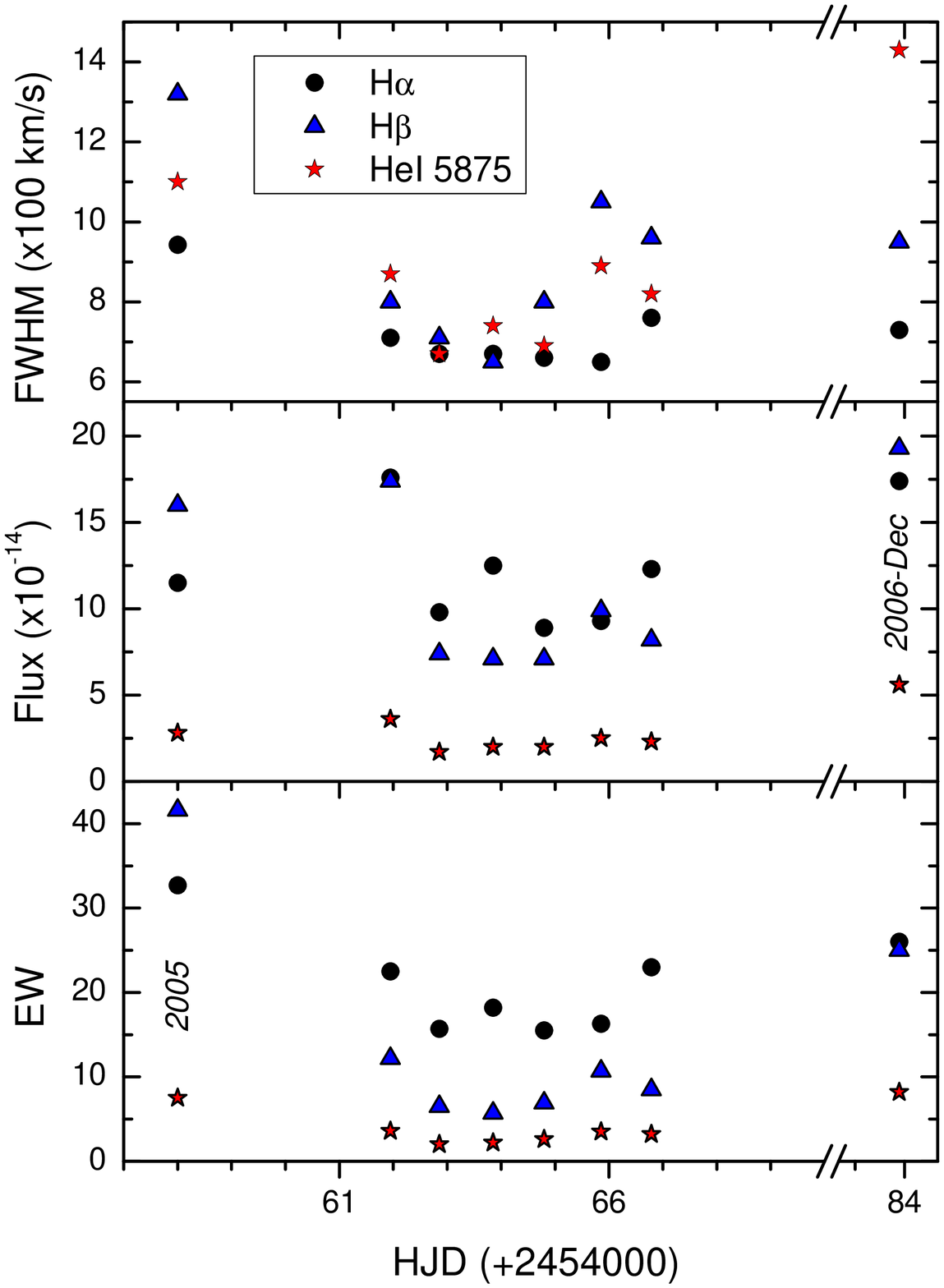}
      \caption{Equivalent Width (EW), line fluxes and Full Widths at Half Maximum (FWHM)
       variations of the \Halpha, \Hbeta\ and \HeI\ $\lambda5875$ emission lines observed in BF~Eri. }
      \label{line_param}
   \end{figure}

  At the beginning of the \textit{2006-Nov} set of observations we found the system to be appreciably
  brighter and bluer than in 2005, being probably in an outburst. We have compared the fluxes in all
  four spectral passbands and found that the brightness of BF~Eri increased by $\sim$1.1 mag in $u'$,
  $\sim$0.8 mag in $b'$, $\sim$0.6 mag in $v'$ and $\sim$0.4 mag in $r'$ colours, relative to the
  \textit{2005-N1} set (Fig.~\ref{colours}).
  During the following nights we observed reddening of the flux distribution and decreasing of the system
  brightness, which almost ceased towards the last two nights. Nonetheless BF~Eri remained to be brighter
  and bluer than in 2005. However when we observed the system again two weeks later (the \textit{2006-Dec} set),
  we found it noticeably weaker than one year previously (Fig.~\ref{colours}).

  The above-described photometric changes we observed throughout the outburst were accompanied by spectral
  changes which however were mostly quantitative rather than qualitative. There was no emergence of broad
  absorption troughs around the emission lines nor dramatic changes in their profiles, as we observed in BZ~UMa
  \citep{Neustroev}. In a quantitative sense the changes became apparent by the decreasing of FWHM, EW and
  flux of the emission lines during the first nights of the \textit{2006-Nov} set (Fig.~\ref{line_param},
  Table~\ref{LineParamTab}). Also of particular interest is tracing the changes of the high excitation lines
  (such as the \HeII\ and \CIII/\NIII\ blend emissions)  which are good tracers of irradiation. Unfortunately,
  due to the weakness of these lines and the poor sensitivity of the CCD at short wavelengths we were able
  only to detect the (probable) presence of \HeII\ $\lambda$4686 in the outburst and no signs of the \CIII/\NIII\
  Bowen blend. Thus, unlike many other dwarf novae which show strengthening of the \HeII\ and \CIII/\NIII\ line
  emissions during an outburst, these lines have remained very weak or even absent in the outburst spectra of
  BF~Eri.

\subsection{Orbital period determination}
\label{PeriodSec}

\begin{figure}
\centering
\includegraphics{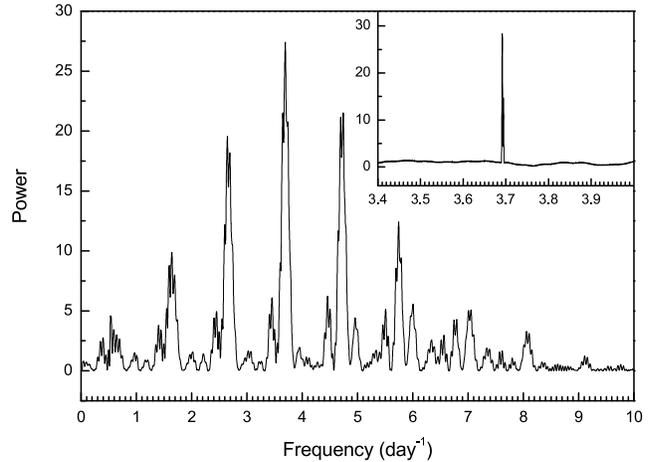}
\caption{Lomb-Scargle power spectrum of the H$\alpha$ radial velocity variations from
the \textit{2006} dataset. The inset shown is blow-up of the most prominent peak of the CLEANed power
spectrum of the combined data.}
\label{power}
\end{figure}

In order to  determine the orbital period of BF~Eri we measured the radial
velocity variations of H$\alpha$ by convolving the observed line profiles
with a single Gaussian of FWHM = 600 km s$^{-1}$. The spectra were continuum-
normalized prior to this analysis. The data were studied for periodicities
using the standard Lomb-Scargle power spectrum analysis \citep{Lomb, Scargle}.
Due to the longer observing span and more homogeneous nature of the data
we initially studied the observations from the \textit{2006} set.
The resulting periodogram shows a strong signal at the 3.6940 day$^{-1}$
frequency (Fig.~\ref{power}), which corresponds to the orbital period of
$P_{orb}=0.2707\pm0.0008$ days.

A more accurate value of the orbital period was determined from the combined data.
We applied the CLEAN procedure \citep{CLEAN} to sort out the alias periods resulting
from the uneven data sampling and daily gaps and obtained a strong peak in the power spectrum
at the 3.69165 day$^{-1}$ frequency, corresponding to the orbital period of
$P_{orb}=0.270881\pm0.000003$ days (Fig.~\ref{power}), consistent with the determination of \citet{BF-Eri}.

%************************  Average Spectra  ************************************

\subsection{The radial velocity of the white dwarf}
\label{radialwd}

    In CVs the most reliable parts of the emission line profile
    for deriving the radial velocity curve are the extreme wings. They
    are presumably formed in the inner parts of the accretion disc and therefore
    should represent the motion of the WD with the highest reliability.
    We measured the radial velocities using the double-Gaussian method
    described by \citet{sch:young} and later refined by \citet{Shafter}.
    This method consists of convolving each spectrum with a pair of Gaussians of
    width $\sigma $ whose centres have a separation of $\Delta $.
    The position at which the intensities through the two Gaussians become equal
    is a measure of the wavelength of the emission line. The measured velocities
    will depend on the choice of $\sigma $ and $\Delta $, and by varying $\Delta $
    different parts of the lines can be sampled. The width of the Gaussians $\sigma $
    is typically set by the resolution of the data.
    In order to test for consistency in the derived velocities and the zero phase, we separately
    used the emission lines \Halpha\ and \Hbeta\ in the \textit{2006-Nov} phase-averaged
    spectra, and the emission line \Halpha\ in
    the phase-averaged spectra from the \textit{2005-N3} set.
    All measurements were made using a Gaussian FWHM of 100 ${\rm km~s^{-1}}$ (and also of
    50 ${\rm km~s^{-1}}$ for the \textit{2006} dataset) and
    different values of the Gaussian separation $\Delta$ ranging from 300
    \kms\ to 2000 \kms\ in steps of 50 \kms,
    following the technique of ``diagnostic diagrams'' \citep{Shafter2}.

    For each value of $\Delta$ we made a non-linear least-squares fit of the derived
    velocities to sinusoids of the form
        \begin{equation}  \label{radvelfit}
          V(\phi,\Delta )=\gamma (\Delta )-K_1(\Delta )\sin \left[ 2\pi \left(
          \phi-\phi_0\left( \Delta \right) \right)\right]
        \end{equation}
    \indent
    where $\gamma$ is the systemic velocity, $K_1$ is the semi-amplitude, $\phi_0$ is
    the phase of inferior conjunction of the secondary star and $\phi$ is the phase
    calculated relative to epoch $T_0=2453675.2670$ for the \textit{2005-N3} set and $T_0=2454061.8112$ for
    the \textit{2006-Nov} set.

\begin{figure}
\centering
\includegraphics[width=8.0cm]{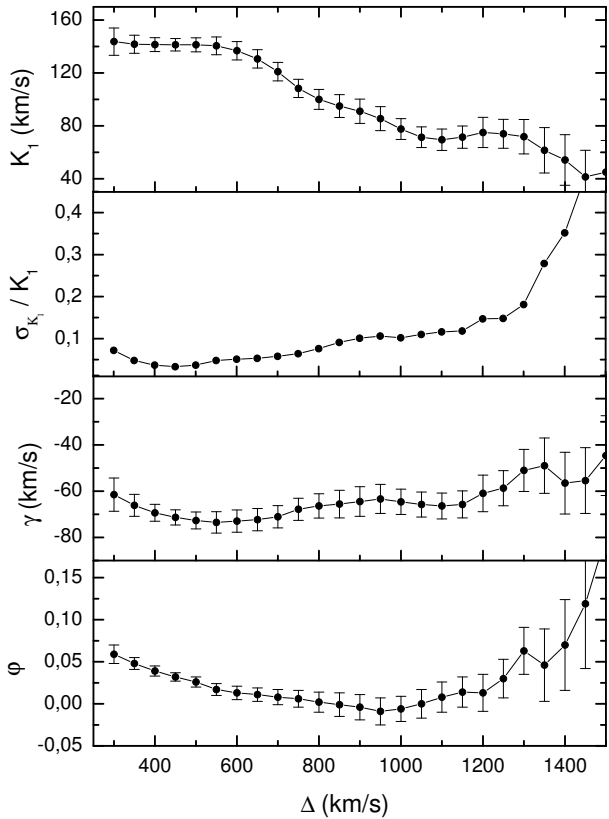}
\caption{The diagnostic diagram for the H$\alpha$ \textit{2006-Nov} data, showing the response of the
fitted orbital elements to the choice of the double-gaussian separation. The best fit is reached with
the gaussian separation of 1150--1250 \kms.}
\label{diagram}
\end{figure}

\begin{figure*}
    \centering
     \includegraphics[width=8.0cm]{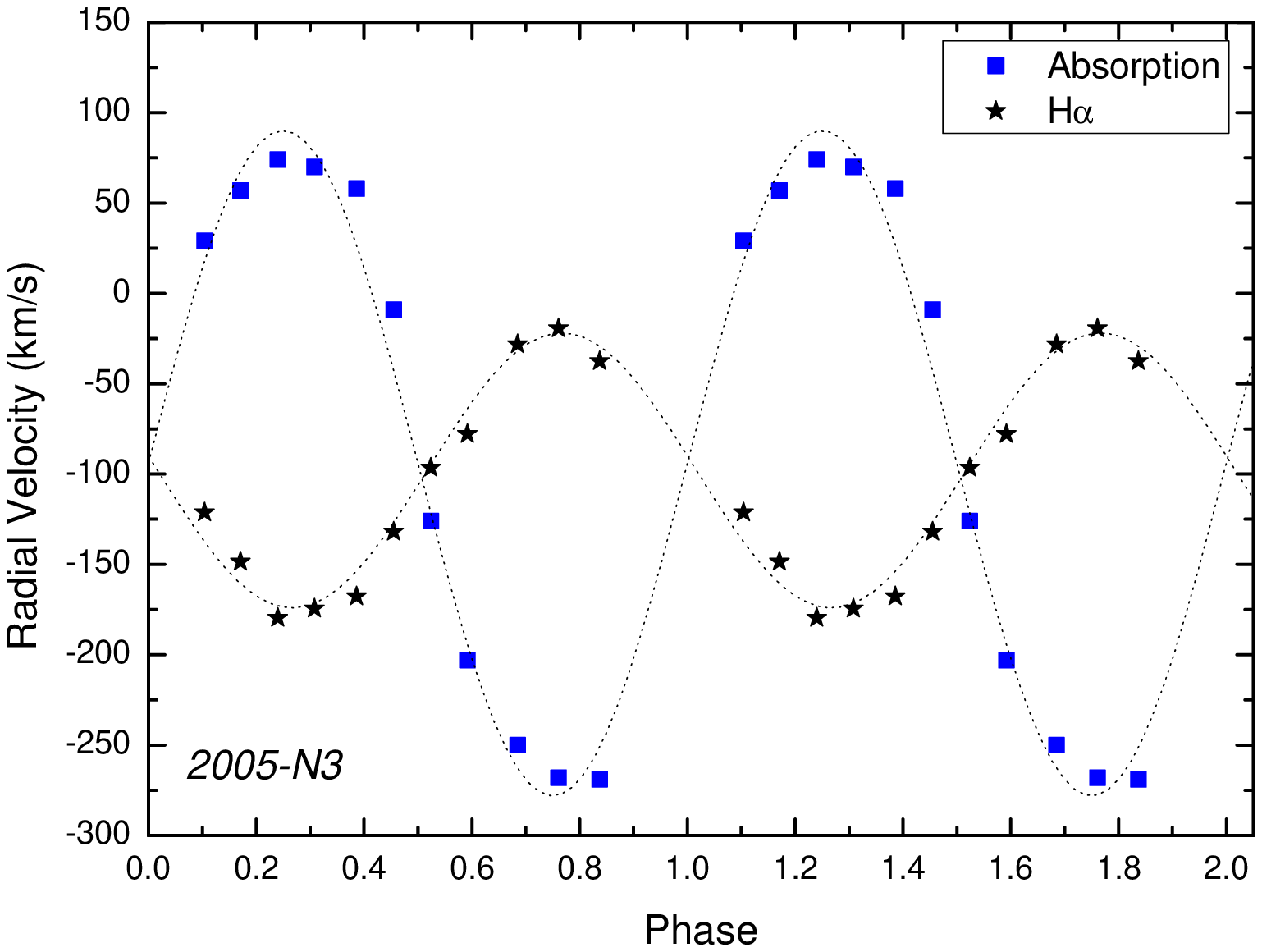}  \includegraphics[width=8.0cm]{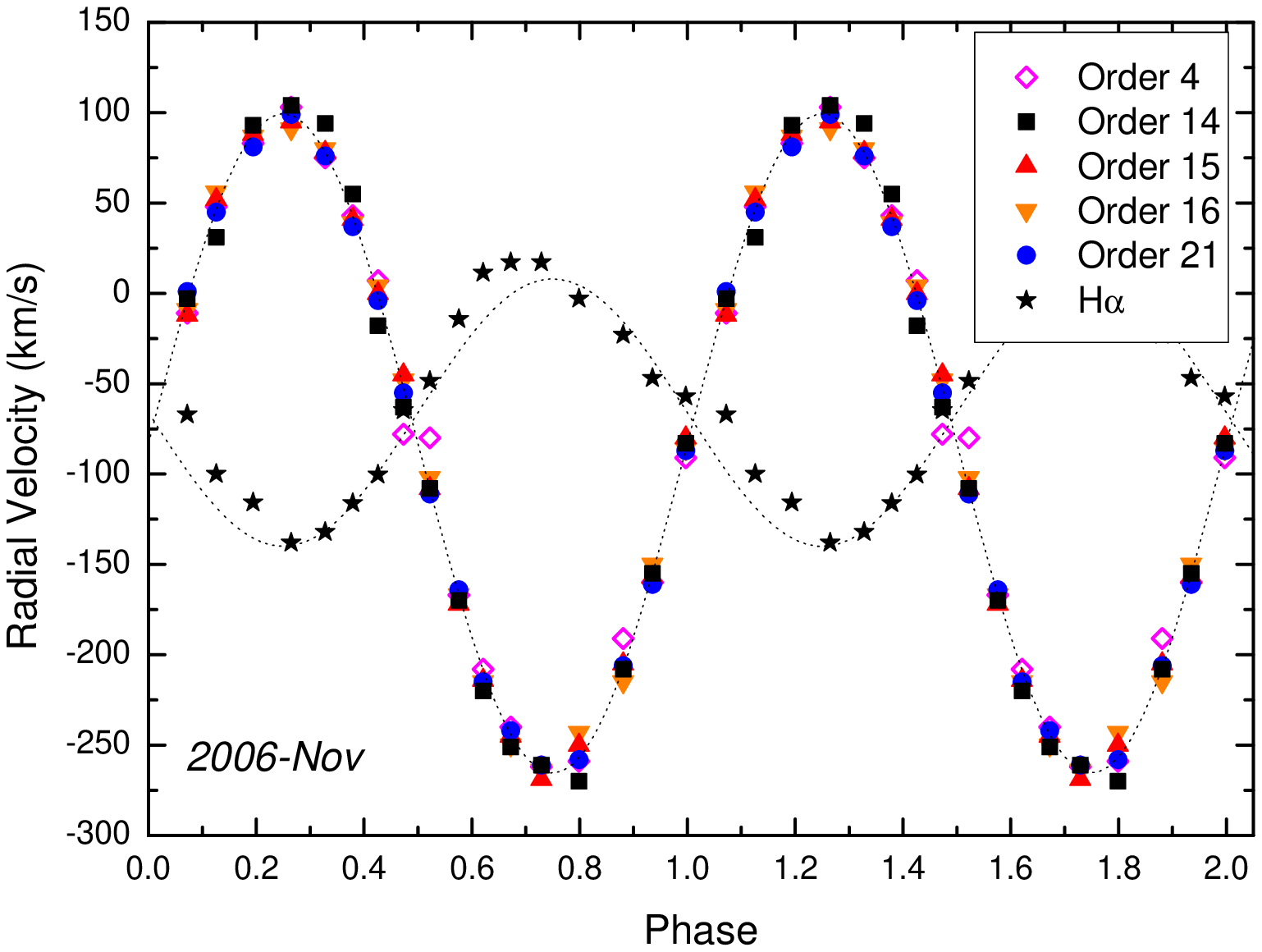}
     \caption{Radial velocities of the white dwarf and the secondary star in BF~Eri folded on the ephemeris from
     Table~\ref{Tab:Syspar}.
     Observations were performed in 2005 November (left panel) and 2006 November (right panel).
     Two cycles are shown for clarity.}
     \label{FigRadVel}
   \end{figure*}

    The resulting ``diagnostic diagram'' for \Halpha\ in the \textit{2006} dataset, with $\sigma=100$ km s$^{-1}$,
    is shown in Fig.~\ref{diagram}. The diagram shows the variations of $K_1$, $\sigma(K_1)/K_1$
    (the fractional error in $K_1$), $\gamma$ and $\phi_0$ with $\Delta $ \citep{Shafter2}.
    The diagrams for other lines and $\sigma$ look the same.
    To derive the orbital elements of the line wings we took the values that
    correspond to the largest separation just before $\sigma(K_1)/K_1$ shows a
    sharp increase \citep{Shafter3}. We find the maximum useful
    separation to be $\Delta \simeq 1100{-}1250$ ${\rm km~s^{-1}}$.
    In Fig.~\ref{diagram} it appears that $\Delta$ can be increased to
    $\sim $1150--1250 km s$^{-1}$ before $\sigma(K)/K$ begins to increase.
    We also note that all the calculated orbital elements are very stable,
    even for the smaller separation $\Delta$, hence there are no difficulties in choosing their values.
    This statement is also true for other ``diagnostic diagrams''.
    We have obtained very consistent results for both the radial velocity semi-amplitudes and
    the $\gamma$-velocities for all of them, and we adopt the mean values $K_1=74\pm3$ \kms\ and
    $\gamma=-91\pm7$ \kms. The measured parameters of the best fitting radial velocity curves
    are summarized in Tables~\ref{TabRadVelEmission} and \ref{Tab:Syspar}. In Fig.~\ref{FigRadVel}
    we show the radial velocity curves for the H$\alpha $ emission line.

    The derived value of the radial velocity semi-amplitude $K_1$ is highly inconsistent with the one found
    by \citet{BF-Eri}. We are unable to explain why their result is so different, as they did not present their
    analysis in detail. On the other hand, we have no reason to doubt our values. A possible reason might be
    poor spectral resolution for their data and as a consequence the wider Gaussians used in the double-Gaussian method.
    According to the ``diagnostic diagram'' (Fig.~\ref{diagram}) we presume that \citet{BF-Eri} examined the
    lower-velocity parts of the \Halpha\ line and consequently obtained a higher value of $K_1$.

%************************  Diagram  ************************************
%************************  Diagram  ************************************

\begin{table}
\caption[]{Elements of the radial velocity curves of BF~Eri derived from emission lines}
\begin{center}
\begin{tabular}{cccc}
\hline\hline
\noalign{\smallskip}
Emission line \& & K$_{1}$ & $\gamma$-velocity $^a$ & $\phi_{0}$$^b$ \\
dataset & (km s$^{-1}$) & (km s$^{-1}$) &  \\
\noalign{\smallskip}
\hline
\noalign{\smallskip}
\Halpha\ - 2005 & 74$\pm$4  &  -98$\pm$4  & 0.013$\pm$0.011 \\
\Halpha\ - 2006 & 74$\pm$4  &  -66$\pm$5  & 0.012$\pm$0.015 \\
\Hbeta\ - 2006  & 77$\pm$10 &  -73$\pm$12 & 0.013$\pm$0.016 \\
%He\,II $\lambda$4686 & 50$\pm$21 & 69$\pm$29 & 2450101.265$\pm$0.004 \\
\noalign{\smallskip}
\hline
\noalign{\smallskip}
\textbf{Mean} & \textbf{74$\pm$3} & \textbf{-91$\pm$7} & \textbf{0.0127$\pm$0.0003} \\
\noalign{\smallskip}
\hline
\noalign{\smallskip}
%Phases were calculated relative to epoch $T_0=2454061.8112$\pm$0.0003$
\end{tabular}
\end{center}
$^a$ The measured $\gamma$-velocities are heliocentric. The mean value was obtained after correction for the solar motion.\\
$^b$ Phases were calculated relative to epoch $T_0=2453675.2670$ for the \textit{2005} set
and $T_0=2454061.8112$ for the \textit{2006} set.
\label{TabRadVelEmission}
\end{table}

\subsection{The radial velocity of the secondary star}
\label{radialsec}

The absorption lines from the secondary star in BF~Eri are visible quite clearly, even in
the individual spectra of the \textit{2005-N3} set, and more easily in the phase-folded spectra of both
the \textit{2005-N3} and \textit{2006-Nov} sets. In order to obtain radial velocities, we used the
following interactive method.

\begin{enumerate}
  \item
    The first step was to measure the radial velocity variations of the absorption line
    \CaI\ $\lambda$6162 (and the \FeI/\CaI/\BaI\ $\lambda$6495 blend for comparison purpose)
    in the phase-folded spectra of the \textit{2005-N3} set. The velocities were measured by convolving
    the observed line profiles with a single Gaussian. The resulting radial velocity curves
    were then fitted with a sinusoid of the form
        \begin{equation}  \label{radvelfit2}
          V(\phi)=\gamma -K_2 \sin \left[ 2\pi \left(\phi-\phi_0 \right) \right]
        \end{equation}
    and preliminary
    values of the radial velocity semi-amplitude $K_2$, the systemic
    velocity $\gamma$ and the phase zero-point $\phi_0$ were calculated.
    The obtained parameters are $K_2=183\pm6$, $\gamma=-94\pm5$ and $\phi_0=0.561\pm0.007$.

  \item   \label{step2}
    Each spectrum in both the \textit{2005-N3} and \textit{2006-Nov} sets was then shifted
    to correct for the orbital motion of the donor star, and the results averaged.
    These averaged spectra of BF~Eri in the rest frame of the secondary star allowed us to
    estimate a preliminary spectral type of the donor star (see Section~\ref{spectype}).
    The averaged spectra of BF~Eri were then cross-correlated with the velocity standards of
    the obtained spectral type and the relative shifts computed and applied.
%\end{enumerate}

\smallskip
    Thus, we obtained the cross-correlation template spectra which were used for measuring the
    absorption-line velocities. This was preferable to cross-correlating with the standard stars for
    the obvious reason that it maximizes the similarity between the template and the individual
    spectra to be cross-correlated, and avoids introducing errors caused by spectral-type mismatch
    between the standard star and the secondary star in BF~Eri, as well as by the incorrectly
    determined rotational broadening of the absorption lines.
\smallskip

%\begin{enumerate}
  \item    \label{step3}
    Next, the phase-folded BF~Eri spectra were cross-correlated with the templates.
    For the \textit{2005-N3} set, the wavelength region 6150--6515 \AA\ was used. To avoid the influence
    of the night-sky lines $\lambda$6300 \AA\ and $\lambda$6363 \AA, some portions of the spectral range
    around these lines were blanked. For the Echelle \textit{2006-Nov} set, the procedure was carried out
    using orders 4, 14, 15, 16 and 21 in the wavelength regions
    4220--4235 \AA, 5160--5235 \AA, 5260--5400 \AA, 5360--5500 \AA\ and 6100--6270 \AA\ %6115--6290
    respectively (Fig.~\ref{echelle_spec}). The solutions obtained by fitting the measured radial velocities with
    sinusoid (\ref{radvelfit2}) are very similar for both sets and all orders.
  \item
    Steps \ref{step2} and \ref{step3} were then repeated and the final results (with almost no changes
    from the previous step) were obtained, so we adopt the averaged values of
    $K_2=182.5\pm0.9$ \kms\ and $\gamma=-93.6\pm0.4$ \kms. These values are consistent with \citet{BF-Eri}.
    We also note that the difference between the phase zero-points obtained from
    the emission and absorption lines is very close to 0.5, as it must be if the derived velocities from those
    lines trace the components' motion.
    In Fig.~\ref{FigRadVel} we show the measured radial velocities together with their sinusoidal fit.
    The measured parameters of the best fitting radial velocity curves are summarized in
    Tables~\ref{TabRadVelAbsorption} and \ref{Tab:Syspar}.
  \item
    Finally, the step \ref{step3} was repeated again, using the standard templates of the best-fitting spectral
    type K3V broadened to the predicted rotational velocity of the mass donor \vsini=78 \kms\
    (see Section~\ref{RotVelSec} and footnote~\ref{rotvelfootnote} for details). No significant difference in
    the measured parameters was found.
\end{enumerate}

\begin{table*}
\caption[]{Elements of the radial velocity curves of BF~Eri derived from absorption lines
by means of cross-correlation.}
%\begin{flushleft}
\begin{tabular}{cccccc}
\hline\hline
\noalign{\smallskip}
Dataset & Spectral & Wavelength   & K$_{2}$ & $\gamma$-velocity $^a$& $\phi_{0}$$^b$ \\
        & Order    & region (\AA) & (km s$^{-1}$) & (km s$^{-1}$) &  \\
\noalign{\smallskip}
\hline
\noalign{\smallskip}
2005 &          & 6150--6515 & 185$\pm$5   & -94$\pm$4 & 0.500$\pm$0.005 \\
2006 & Order 04 & 4220--4235 & 180$\pm$4   & -80$\pm$3 & 0.502$\pm$0.003 \\
2006 & Order 14 & 5160--5235 & 187$\pm$3   & -83$\pm$2 & 0.499$\pm$0.002 \\
2006 & Order 15 & 5260--5400 & 183$\pm$2   & -81$\pm$1 & 0.499$\pm$0.001 \\
2006 & Order 16 & 5360--5500 & 182$\pm$2   & -81$\pm$2 & 0.500$\pm$0.002 \\
2006 & Order 21 & 6100--6270 & 181$\pm$2   & -82$\pm$1 & 0.499$\pm$0.001 \\
\noalign{\smallskip}
\hline
\noalign{\smallskip}
\textbf{Mean} & && \textbf{182.5$\pm$0.9} & \textbf{-93.6$\pm$0.4} & \textbf{0.500$\pm$0.001} \\
\noalign{\smallskip}
\hline
\noalign{\smallskip}
%Phases were calculated relative to epoch $T_0=2454061.8112$\pm$0.0003$
\end{tabular}\\
%\end{flushleft}
$^a$ The measured $\gamma$-velocities are heliocentric. The mean value was obtained after correction for the solar motion.\\
$^b$ Phases were calculated relative to epoch $T_0=2453675.2670$ for the \emph{2005} set
and $T_0=2454061.8112$ for the \emph{2006} set.
\label{TabRadVelAbsorption}
\end{table*}

\subsection{Spectral type of the secondary star}
\label{spectype}
  The best absorption lines for spectral classification in normal stars, which are
  independent of chemical abundances are the line ratios of the \FeI\ lines $\lambda 4250$, $\lambda 4260$,
  $\lambda 4271$ to the \CrI\ lines $\lambda 4254$ and $\lambda 4274$, as well as the strength of
  the \CaI\ $\lambda$4226 line \citep{Keenan-McNeil, Echevarria}.
  The \CrI\ lines increase in strength with respect to the \FeI\
  lines with spectral type. On the other hand the intensity of the \CaI\ line also increases steadily as
  the spectral type advances (in fact this line has a saturation effect for late K and M stars).
  Unfortunately, due to the poor sensitivity of the CCD at short wavelengths the data were too noisy
  in this wavelength region, so we were able to use only the width of the \CaI\ $\lambda$4226 line as
  a spectral indicator. In accordance with this a spectral type of the secondary star in BF~Eri can
  be initially estimated as K3--K4.

  To verify this result, we also analyzed the appearance of the absorption lines in those
  spectral orders where they are visible better. Purely by inspection, we have found several reliable
  indicators to be useful in constraining the spectral type. Among them are pairs of the lines/blends \FeI\ \lam 5227.2
  and \CrI /\FeI\ \lam\lam 5206+5208 (order 14), \FeI\ \lam 5434.5 and \FeI\ \lam 5424.1 (order 16),
  \CaI\ \lam 6122.2 and \FeI\ \lam 6137 (order 21), \CaI\ \lam 6439.1 and \FeI\ \lam\lam 6400.0+6400.3 (order 22).
  The relative depths of all these couples change rapidly over the spectral type range we are interested in here.
  Analysis of these absorption features supports a K2--K3 spectral classification for the donor star in BF~Eri.

  Taking into account all the estimations we adopt the final value for the spectral type of
  the secondary to be K3$\pm$0.5, consistent with the determination of \citet{BF-Eri}.

\subsection{Rotational velocity of the secondary star}
\label{RotVelSec}

  The absorption lines visible in the averaged and orbitally corrected spectra are obviously broadened
  due to orbital smearing during exposure and rotation of the secondary star. The observed rotational velocity
  of the secondary could provide an independent determination of the mass ratio of the binary. We have attempted
  to estimate this parameter by artificially broadening the template star spectra which are assumed to have low
  \vsini, and fitting them to the orbital velocity corrected BF~Eri spectra until the lowest residual was obtained.
  We have restricted this analysis to the spectra acquired with the Echelle spectrograph because of their
  higher spectral resolution. We have selected from them those spectral orders which exhibit the greater set
  of the strong absorption-line features (orders 14, 15, 16 and 21).
  All details of the used \vsini\ measuring technique can be found in \citet{Smith1998}
  [see also \cite{Marsh1994} and \cite{North2000}].

  Unfortunately, despite the efforts expended we have failed to derive consistent value for the rotational
  velocity. Instead we obtain a quite broad range of \vsini\ values of 70--100 \kms, that is
  unacceptable for the determination of the system parameters\footnote{\label{rotvelfootnote}See corresponding discussion
  in Section~\ref{SysParDiscSect}. Note that using the system parameters of BF~Eri, obtained in
  Section~\ref{SysParSec} (Table~\ref{Tab:Syspar}), we get the estimated value of \vsini\ to be $\sim78$ \kms,
  consistent with observations.}. We have additionally examined
  the spectra, varying the wavelength ranges and even cutting the separate absorption lines, but were
  unable to get a more certain value of \vsini. We explain our failure by the insufficient quality of the used spectra
  as even the averaged spectra are quite noisy. We still need more observations with high spectral resolution and
  better time resolution to fully solve this problem.

  Although we could not determine the rotational velocity, we were able to estimate the fractional contribution
  of the secondary star to the total light. We have found that the secondary contributes about 11 per cent of the
  total light in order 14.

   \begin{figure*}
    \includegraphics[width=5.6cm]{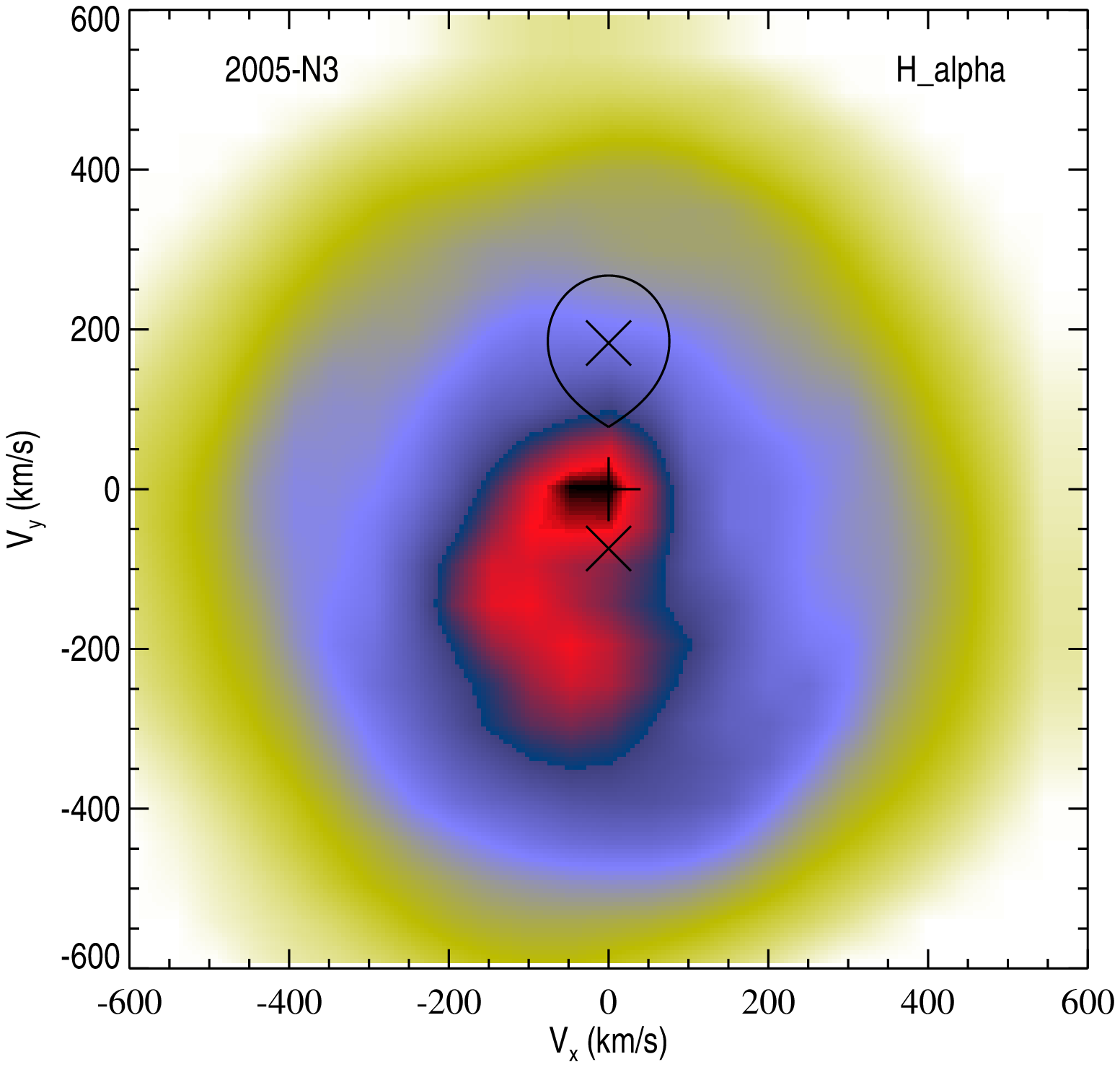}
    \includegraphics[width=5.6cm]{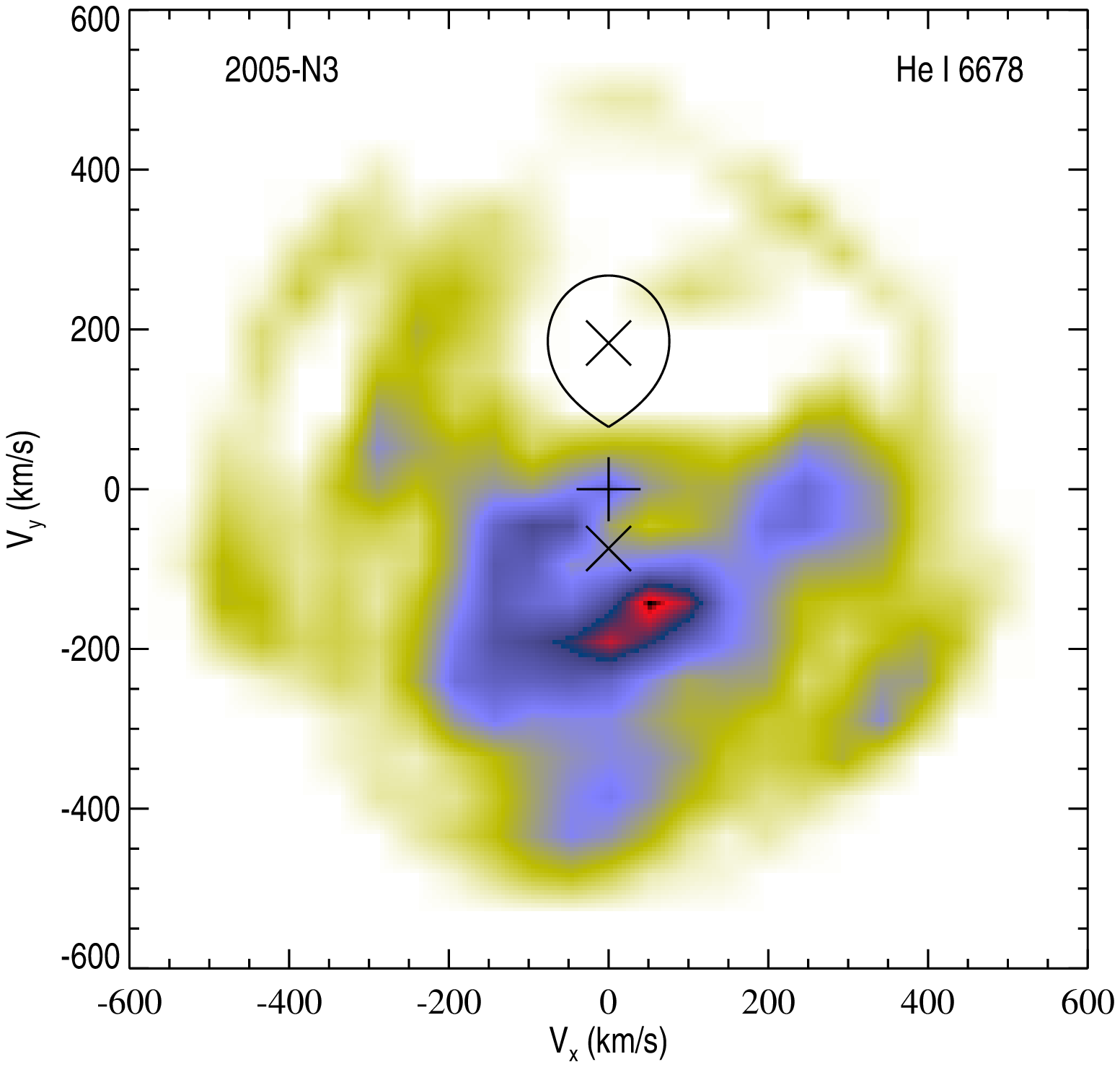}
    \includegraphics[width=5.6cm]{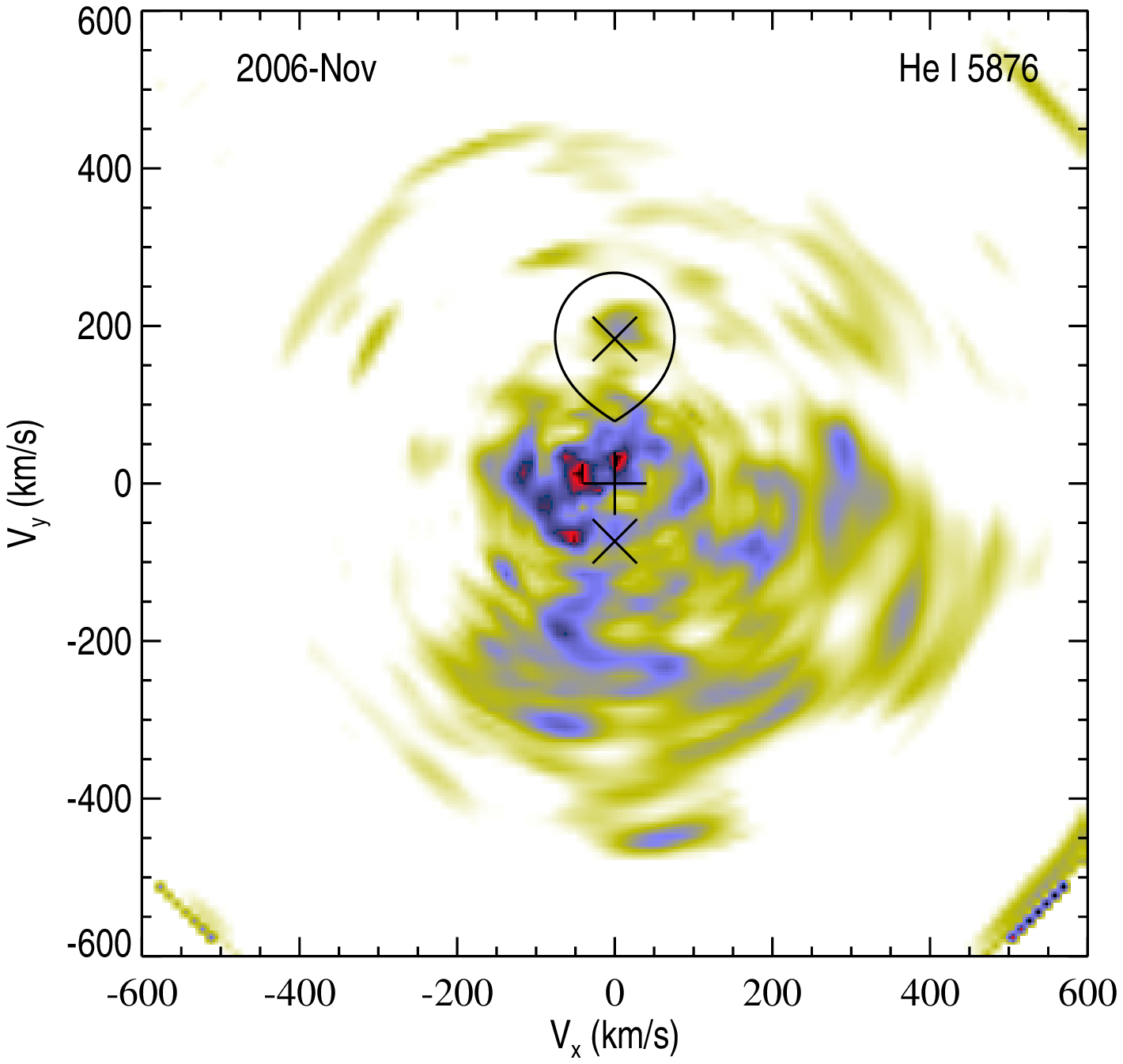}\\
    \includegraphics[width=2.7cm]{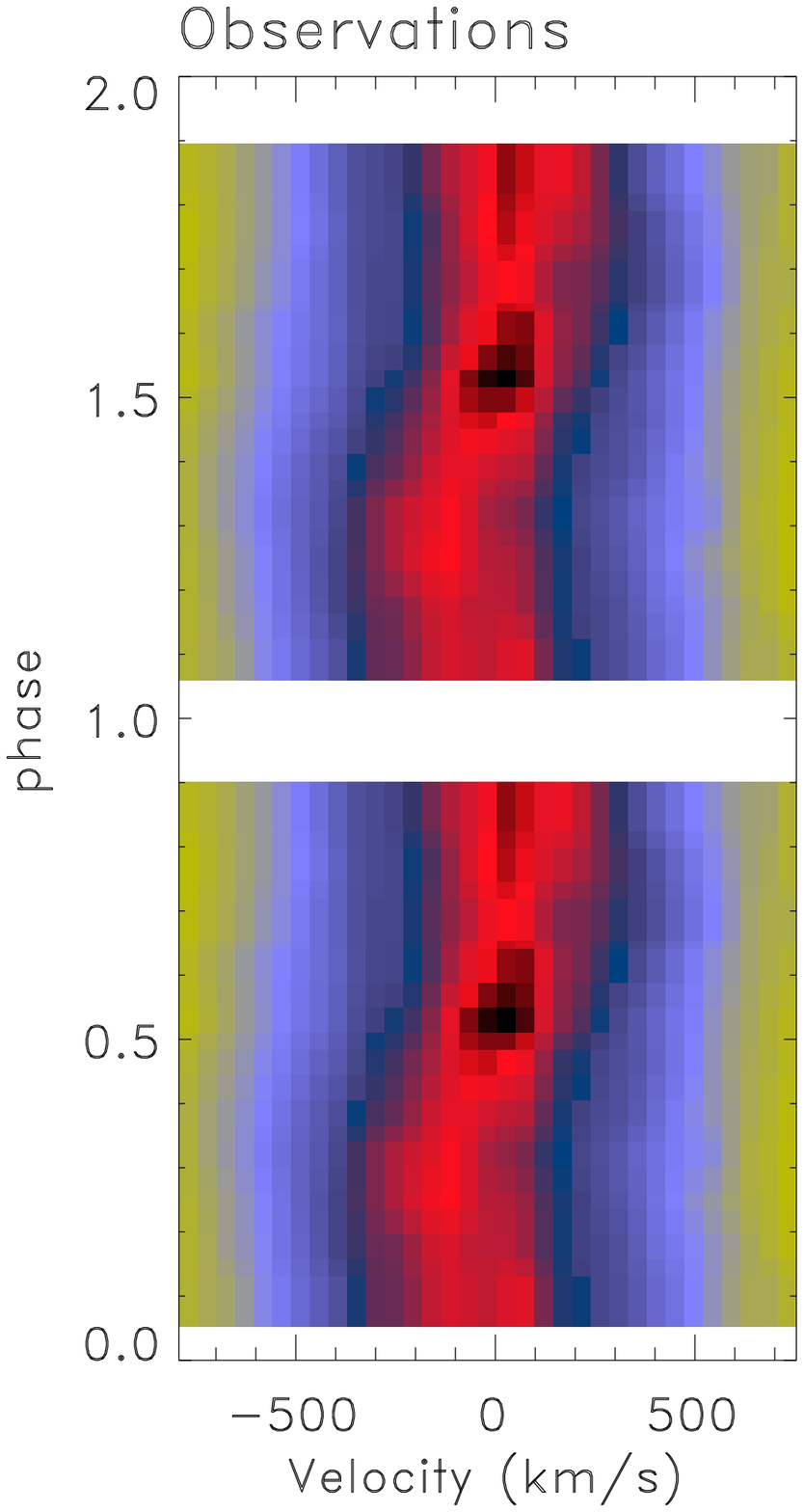}
    \includegraphics[width=2.7cm]{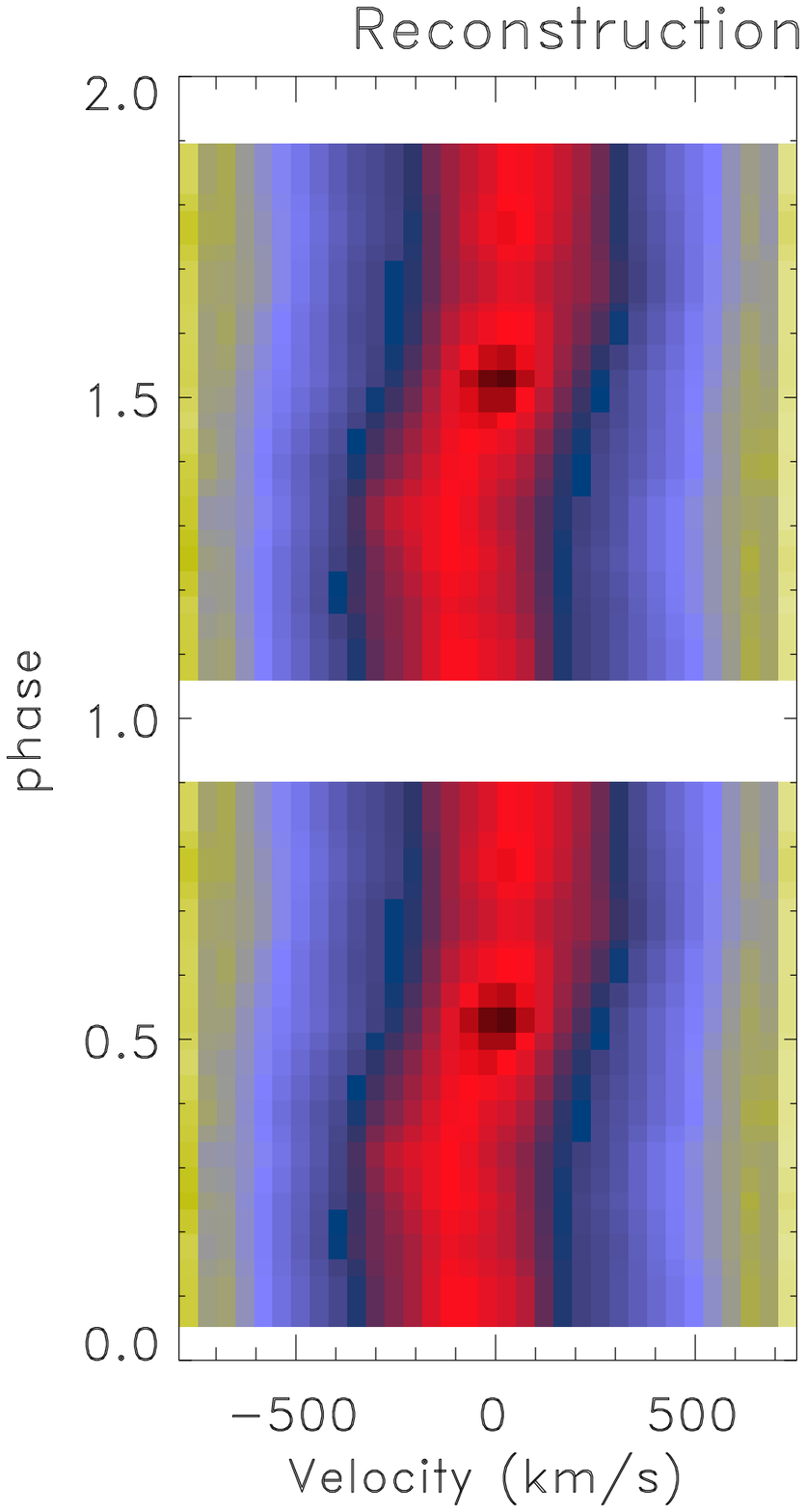}
    \includegraphics[width=2.7cm]{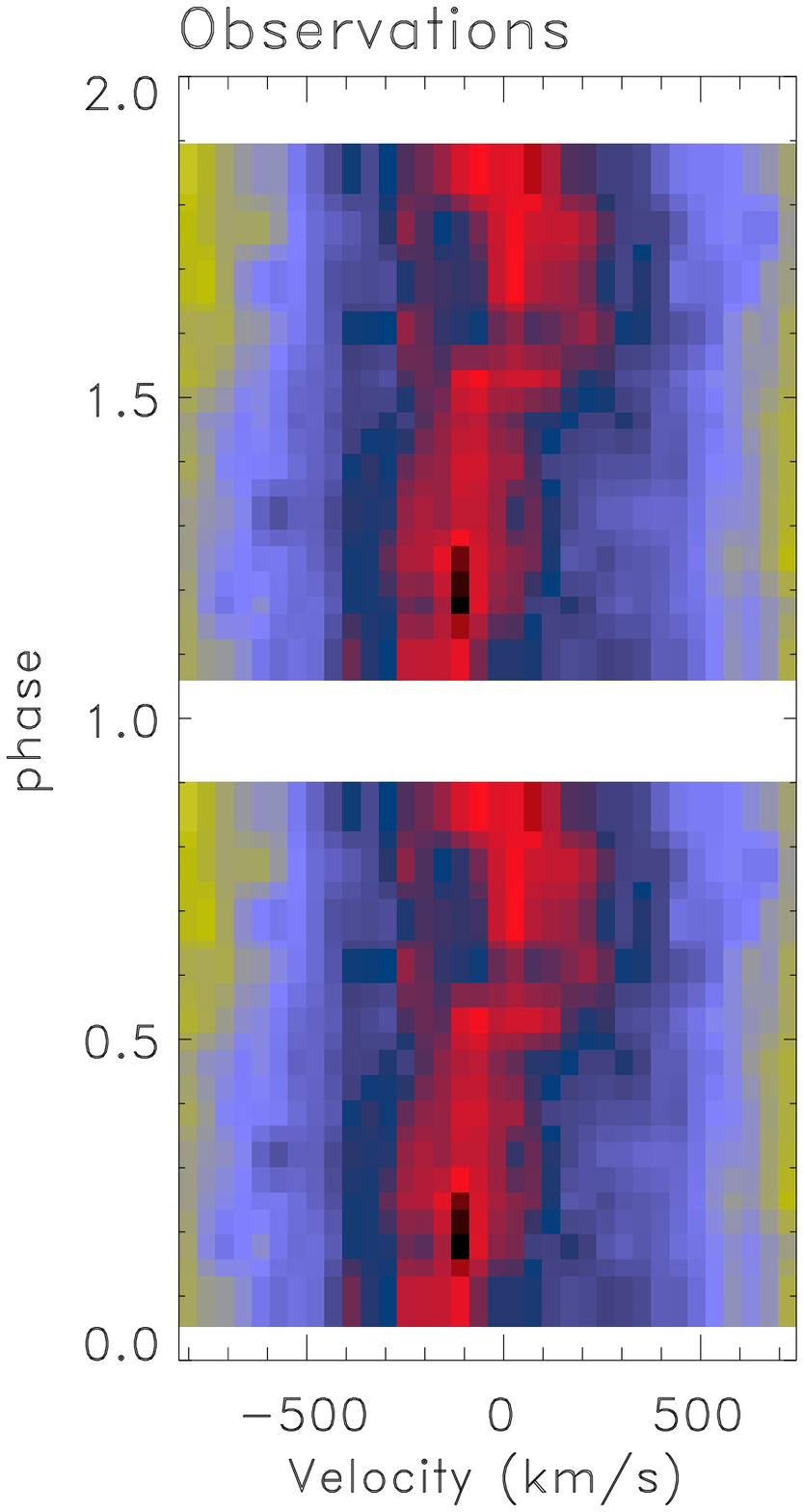}
    \includegraphics[width=2.7cm]{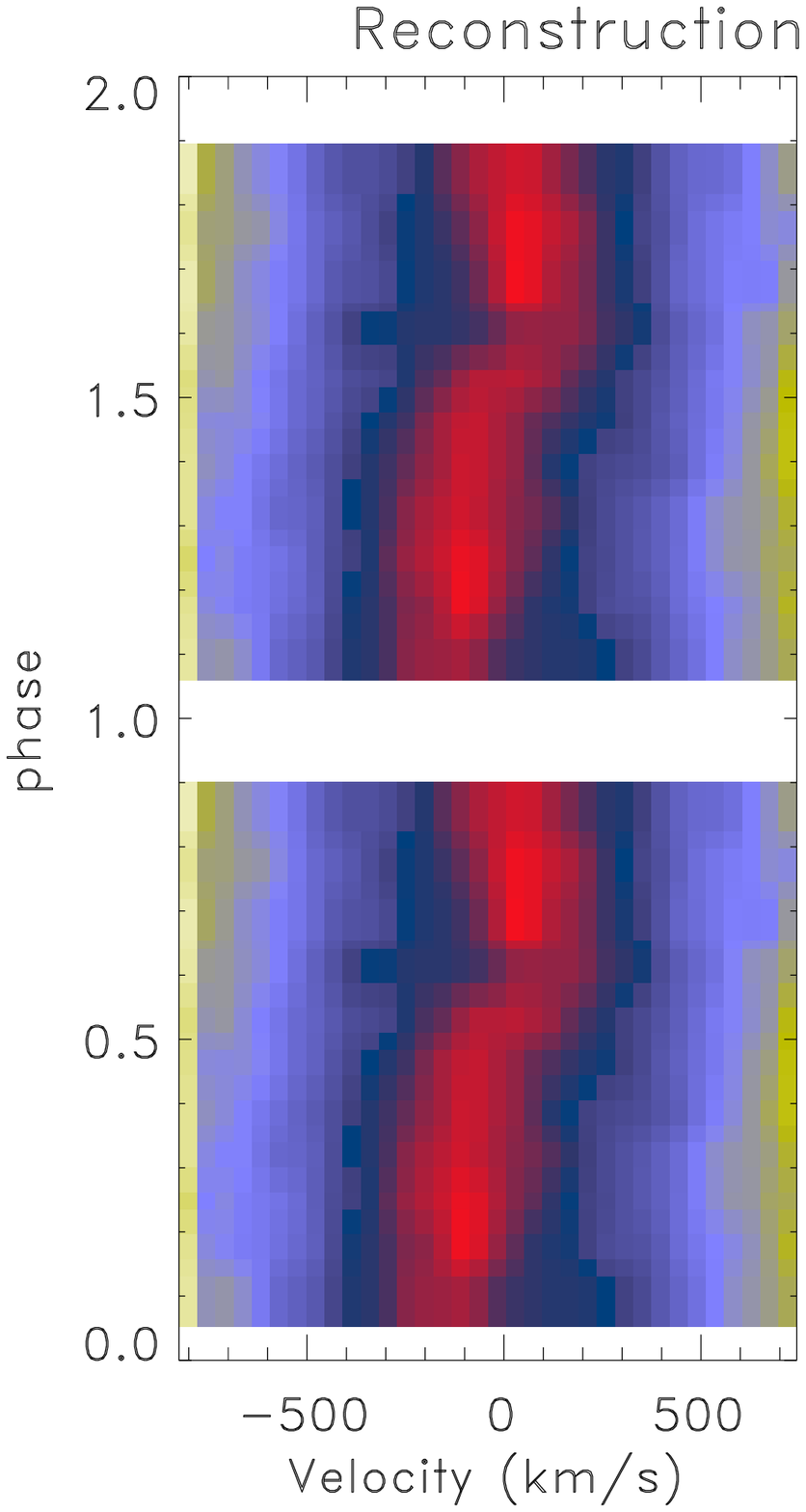}
    \includegraphics[width=2.7cm]{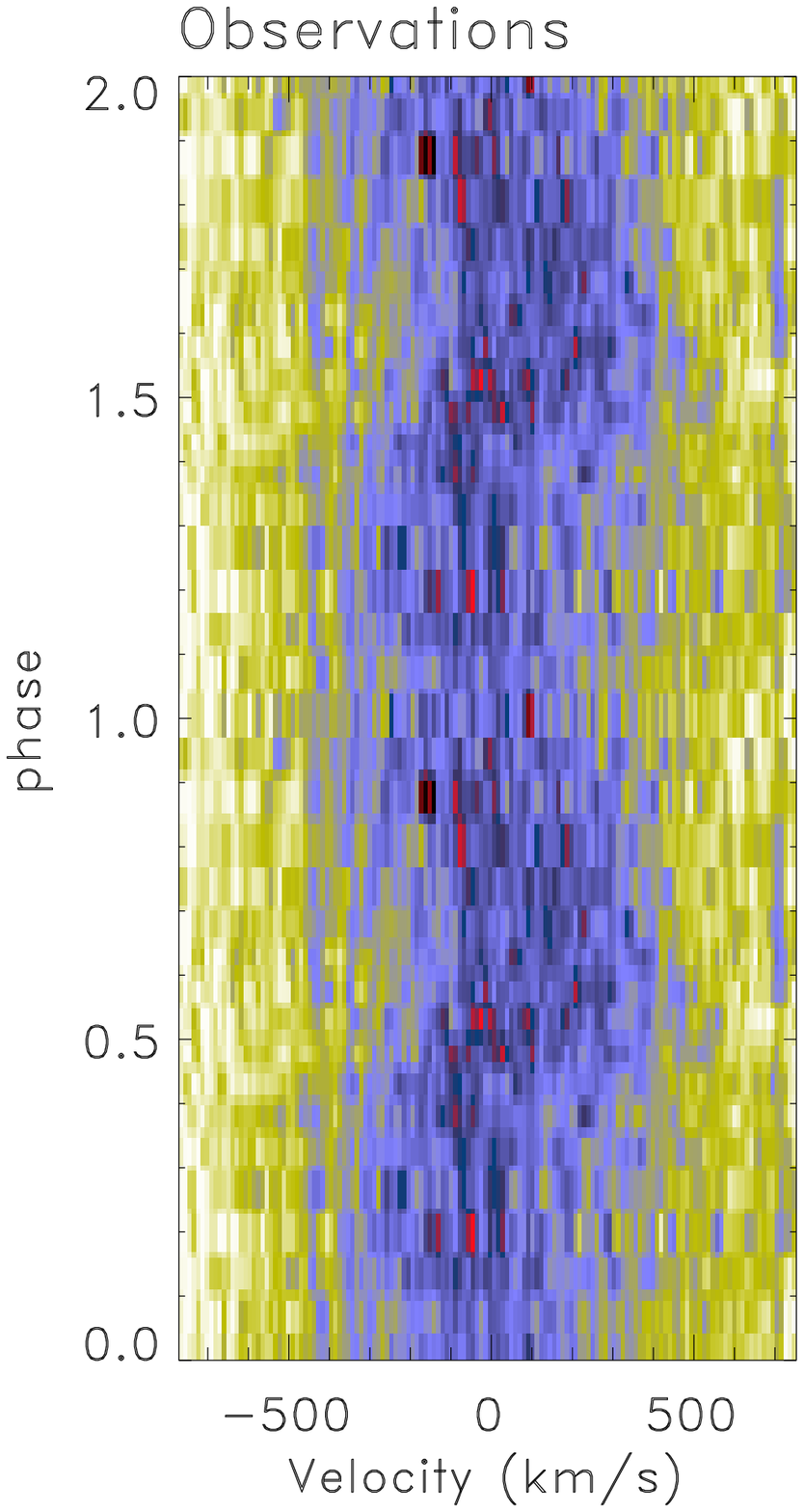}
    \includegraphics[width=2.7cm]{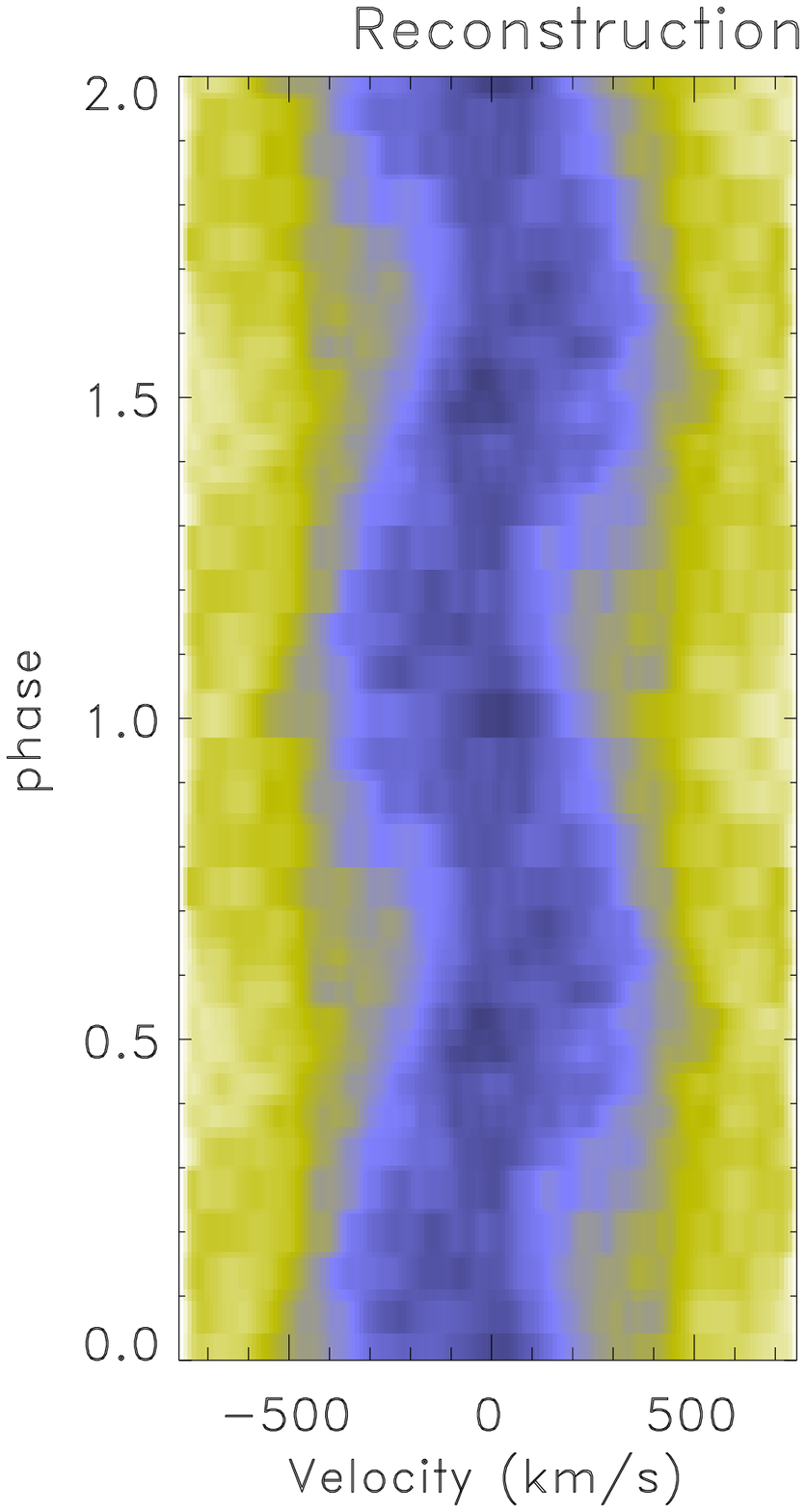}\\
    \bigskip \bigskip
    \includegraphics[width=5.6cm]{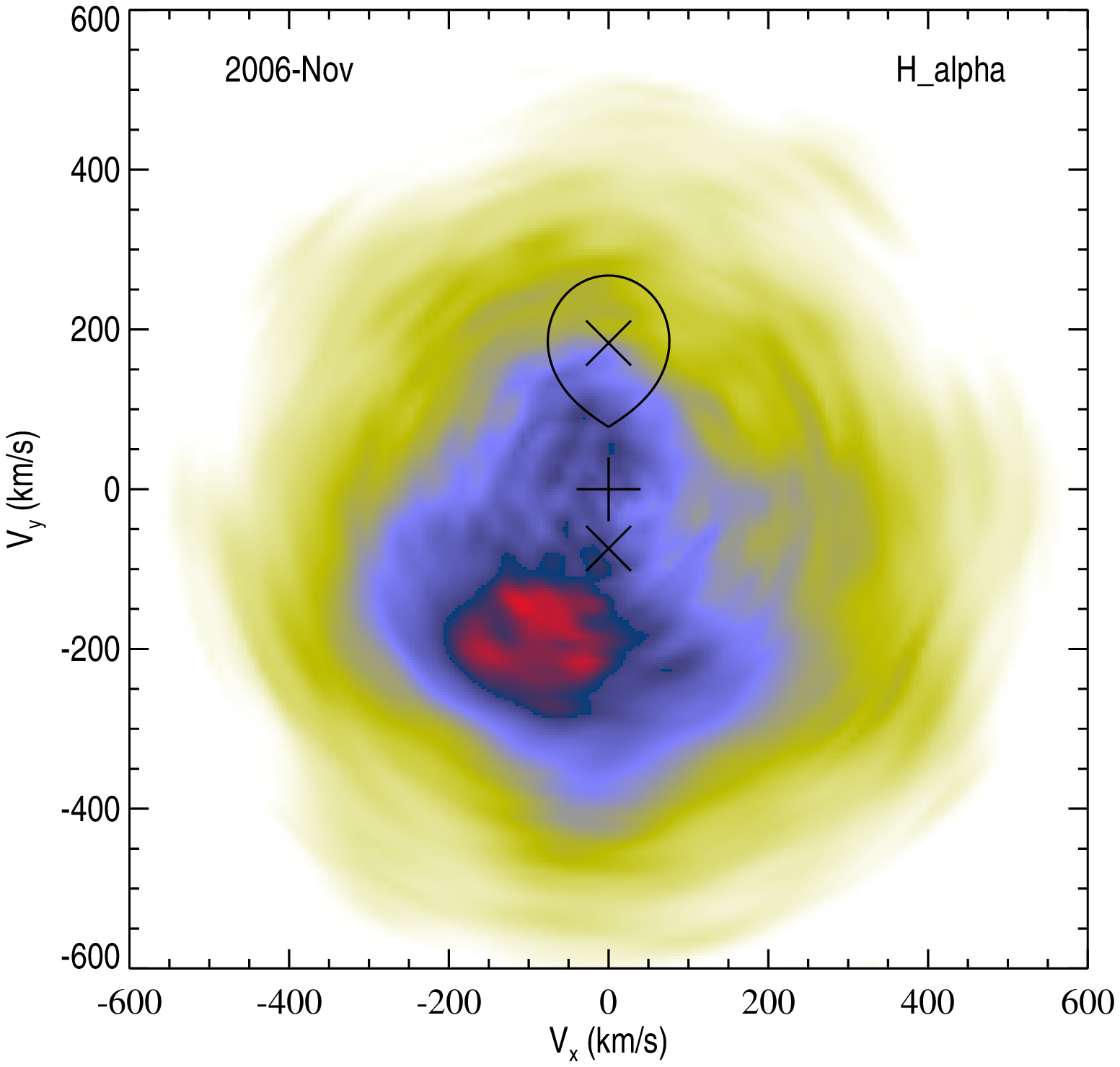}
    \includegraphics[width=5.6cm]{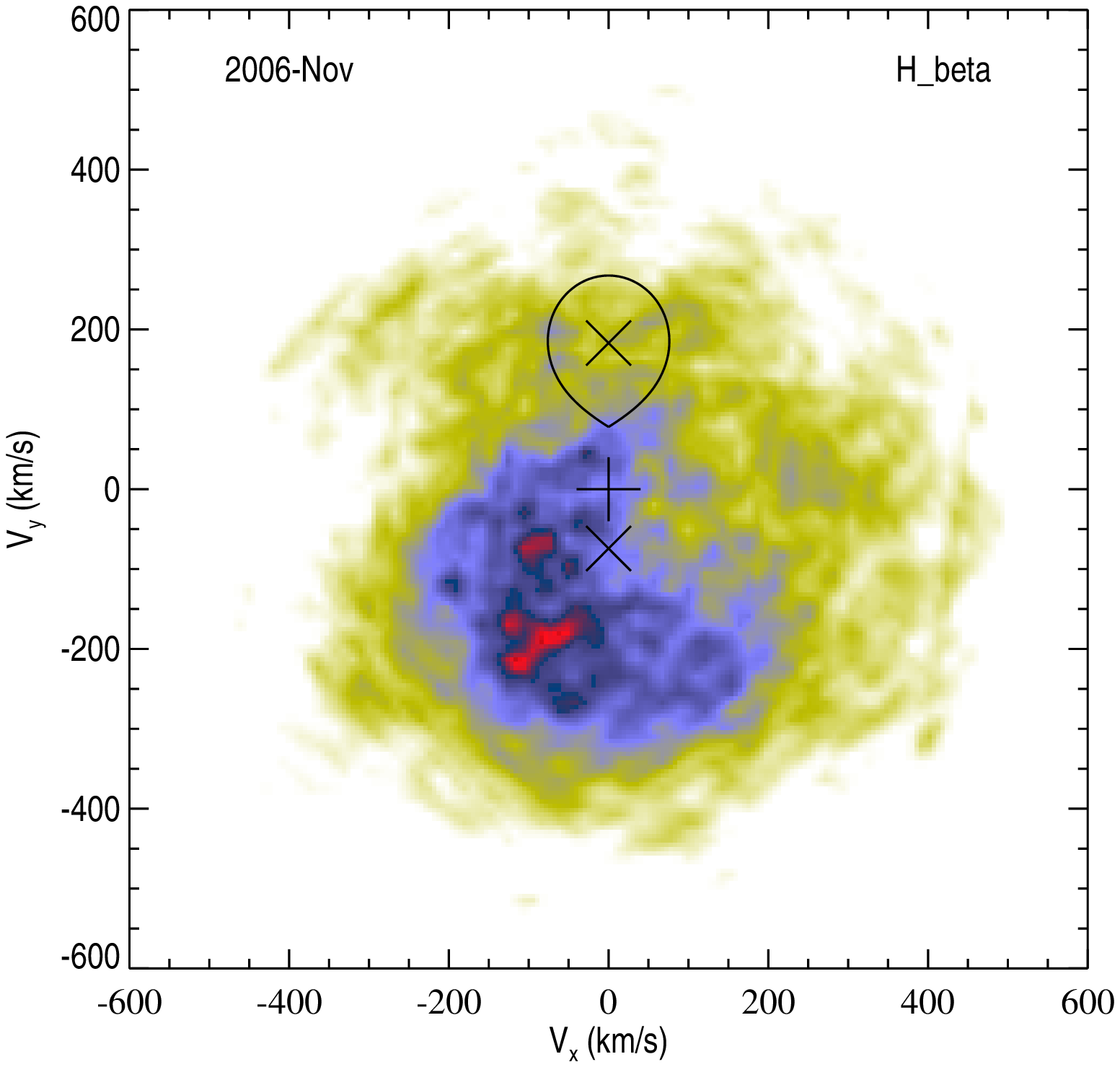}
    \includegraphics[width=5.6cm]{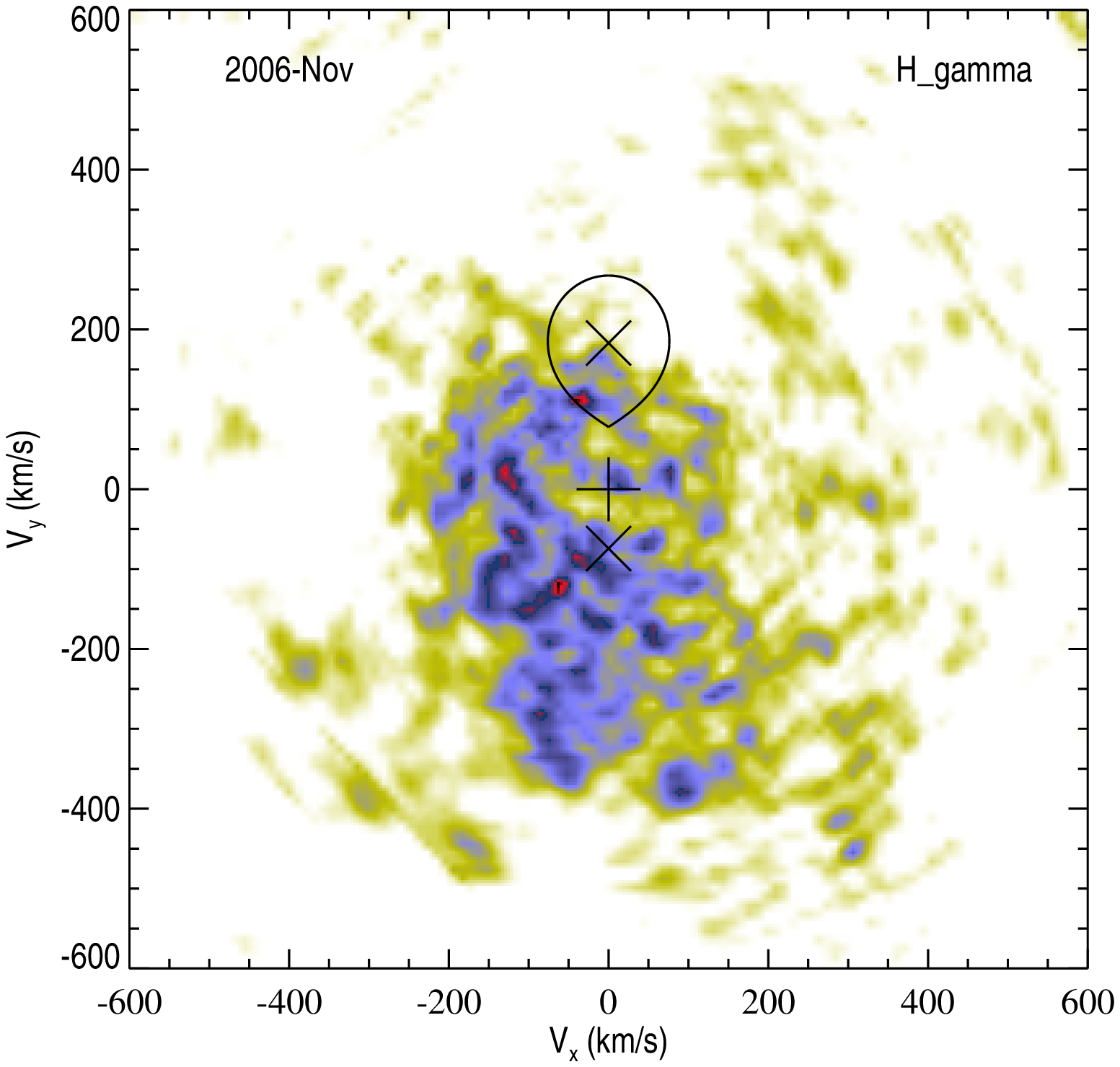}\\
    \includegraphics[width=2.7cm]{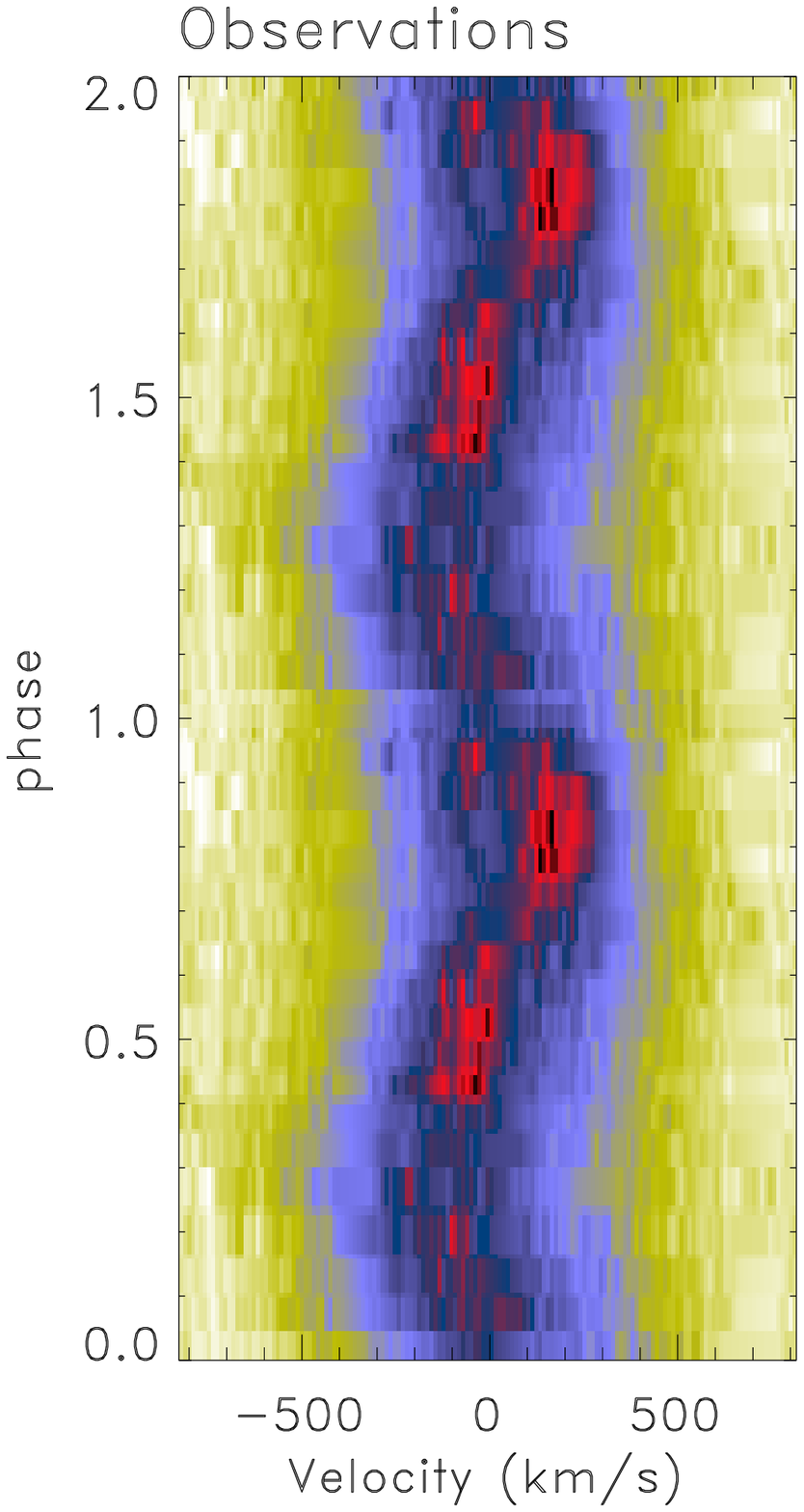}
    \includegraphics[width=2.7cm]{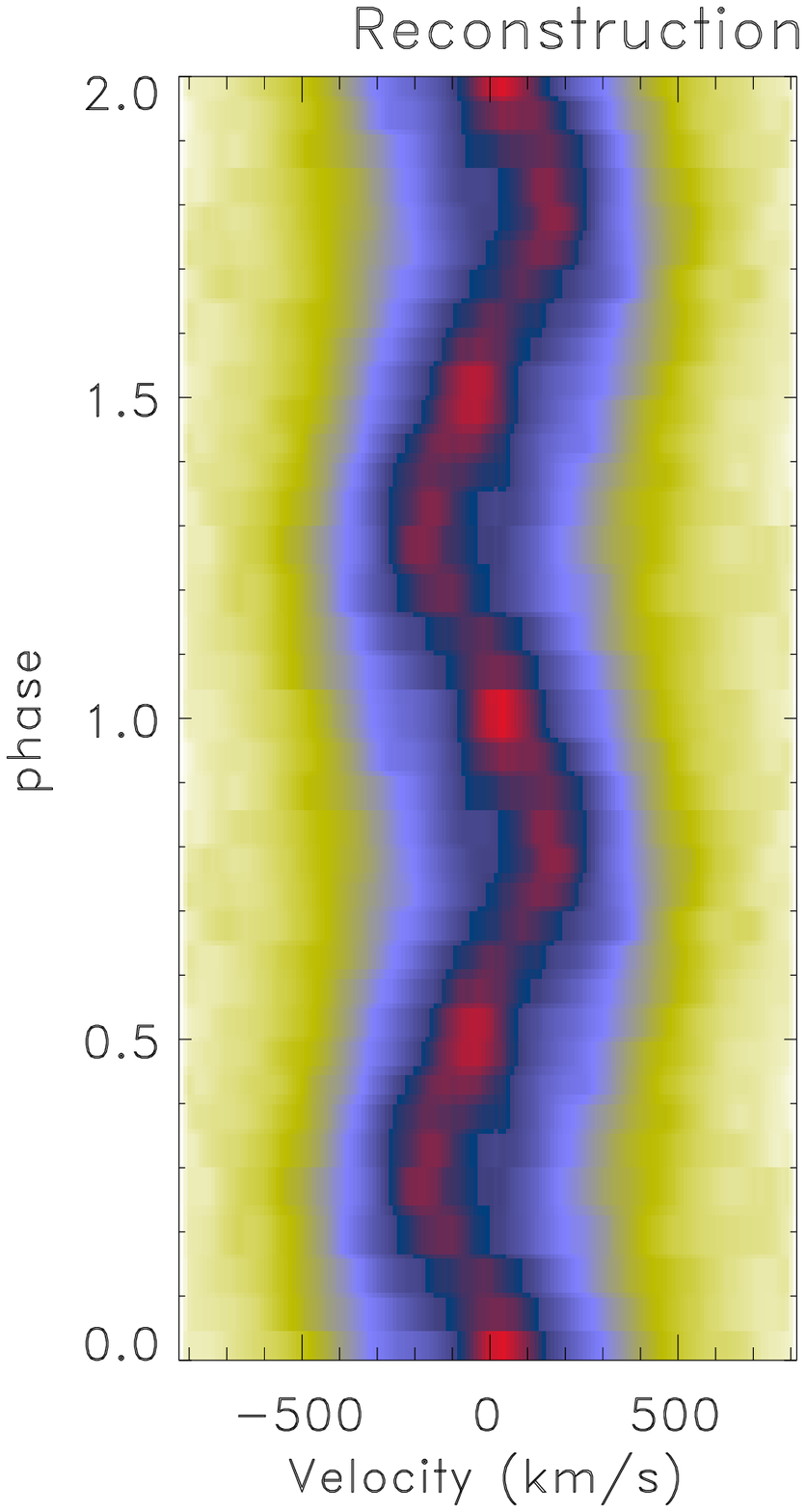}
    \includegraphics[width=2.7cm]{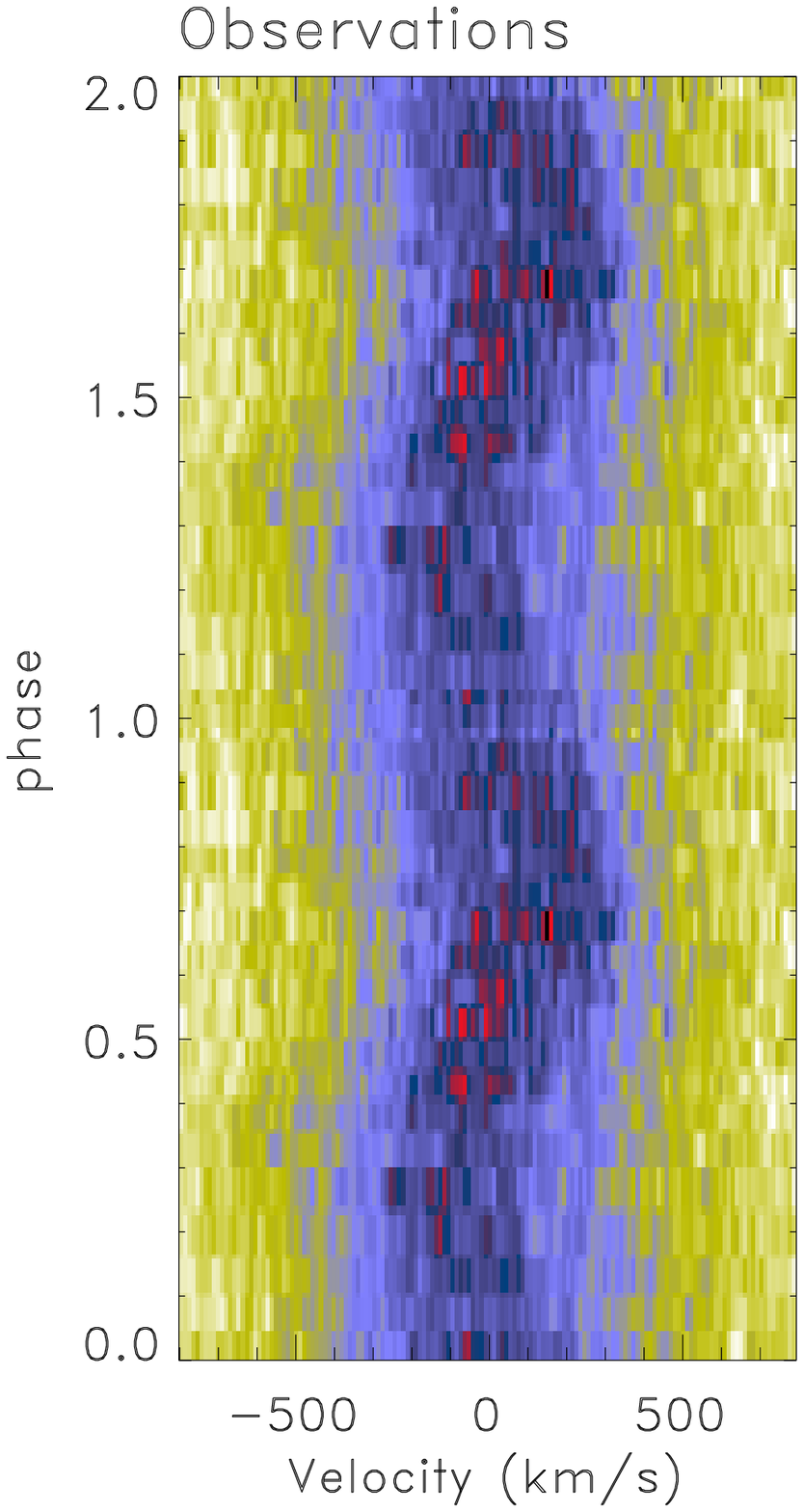}
    \includegraphics[width=2.7cm]{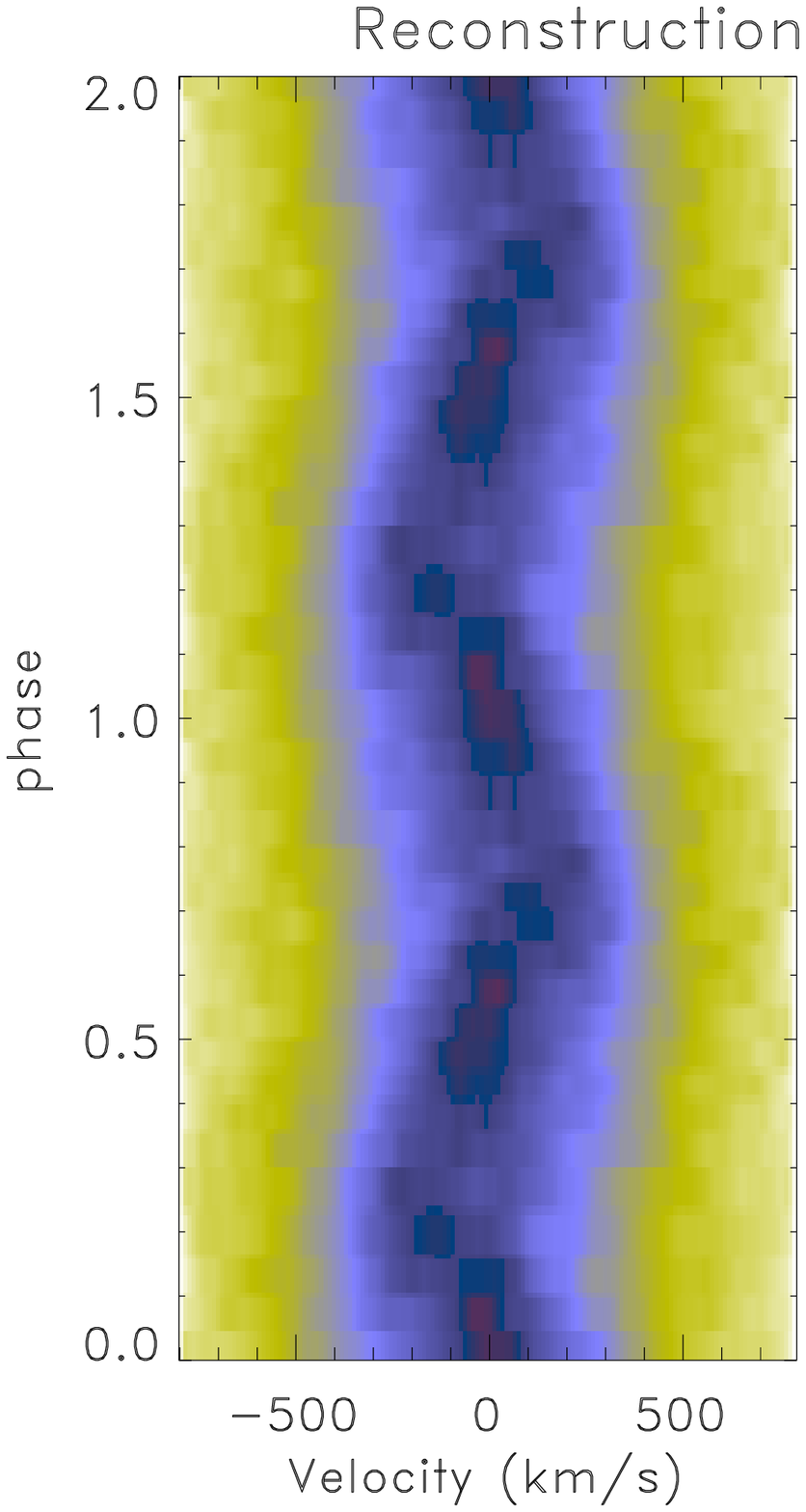}
    \includegraphics[width=2.7cm]{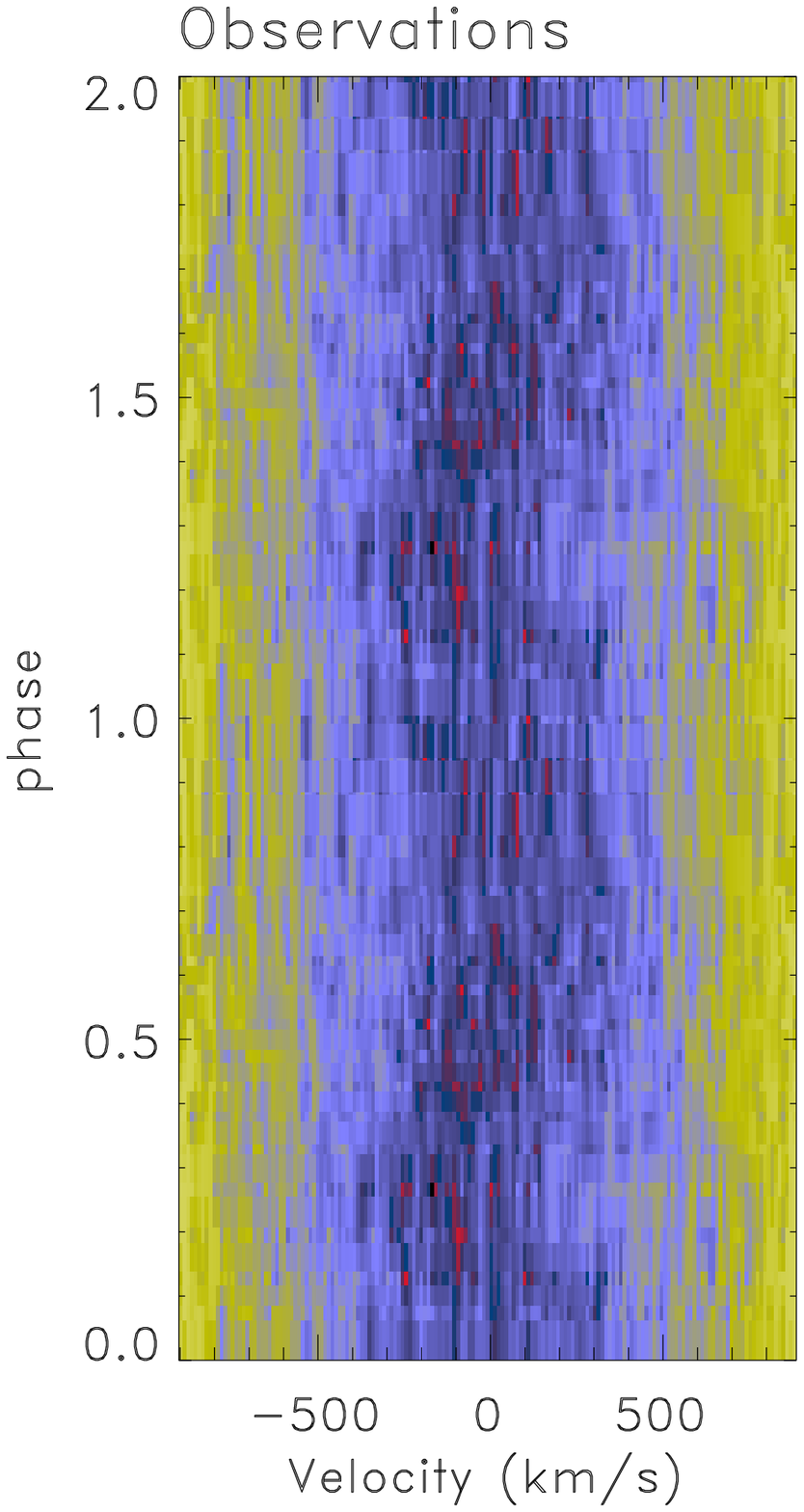}
    \includegraphics[width=2.7cm]{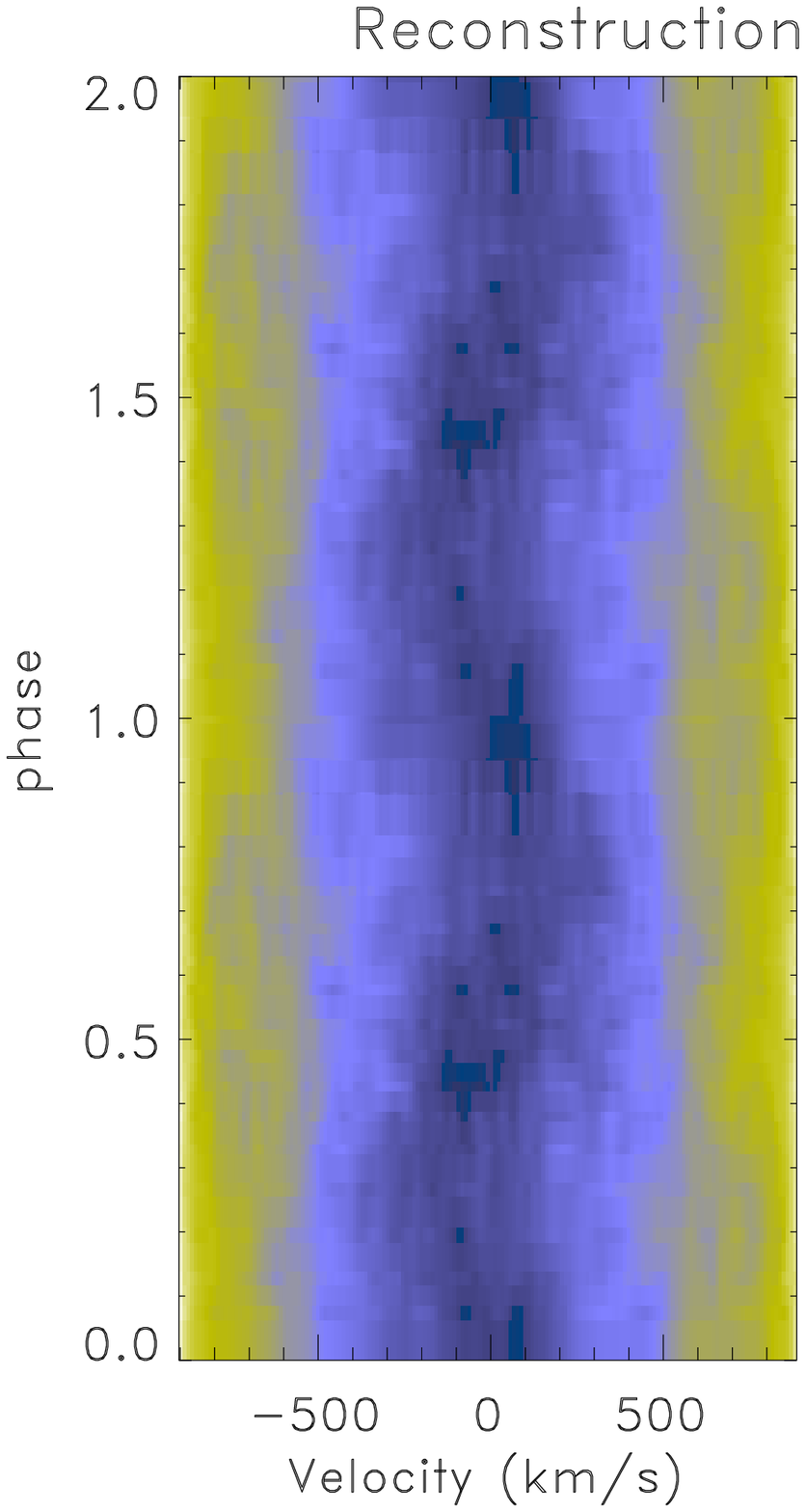}
    \caption{Doppler tomography for the \Halpha\ and \HeI~$\lambda$6678 emission lines from the \textit{2005-N3} set
    and for the \HeI~$\lambda$5876 line from the \textit{2006-Nov} set (in the upper half of Figure), and
    for \Halpha, \Hbeta\ and \Hgamma\ from the \textit{2006-Nov} set of observations (in the bottom half of Figure).
    For each line the observed and reconstructed trailed spectra (bottom) and corresponding Doppler maps (top) are shown.}
    \label{dopmaps}
   \end{figure*}

\section{Doppler Tomography}
\label{DopMapSec}

  The orbital variation of the emission line profiles indicates a non-uniform
  structure for the accretion disc.
  In order to study the emission structure of BF~Eri we have used Doppler tomography.
  Full technical details of the method are given by \citet{Marsh-Horne} and \citet{marsh2001}.
  Examples of the application of Doppler tomography to real data are given by \citet{marsh2001}.

  Figure~\ref{dopmaps} shows the tomograms of the \Halpha\ and \HeI~$\lambda$6678 emissions from the \textit{2005-N3}
  set and of the \Halpha, \Hbeta, \Hgamma\ and \HeI~$\lambda$5876 emissions from the \textit{2006-Nov} set
  of observations, computed using the code developed by \citet{Spruit}. These figures also show trailed
  spectra in phase space and their corresponding reconstructed counterparts.
  A help in interpreting Doppler maps are additional inserted plots which mark the positions
  of the WD (lower cross), the center of mass of the binary (middle cross) and
  the Roche lobe of the secondary star (upper bubble with the cross). The Roche lobe of the secondary
  has been plotted using the system parameters, derived in Section~\ref{SysParSec}.

  Due to the non-double-peaked emission line profiles of BF~Eri, we did not expect a Doppler map to have
  an annulus of emission centered on the velocity of the WD, and the observed tomograms do not show it.
  Instead of this all the maps display a similar, very nonuniform distribution of the emission, the bulk of
  which is located in the lower-left quadrant of the tomograms, at lower velocities than the
  predicted outer disc velocities for an accretion disc with a radius $R_{L_1}$.
  The appearance of the \textit{2005-N3} \Halpha\ map is similar to the hydrogen line maps from
  the \textit{2006-Nov} set but a most prominent contribution
  of the emission here is from the area around the center of mass of the binary system.
  Note also that BF~Eri does not show any emission on the hemisphere of the donor star facing
  the WD and boundary layer.

  The interpretation of the emission structure in BF~Eri is ambiguous.
  All the detected emission sources are located far from the region of interaction between the stream
  and the disc particles. This 'reversed bright spot' phenomenon can perhaps be explained by a
  gas stream which passes above the disc and hits its back \citep{Hellier-Robinson}, or alternatively, by the disc
  thickening in resonating locations.

\section{The binary system parameters}
\label{SysParSec}

   It is impossible to determine accurate system parameters for a non-eclipsing binary system like BF~Eri.
   However, we can constrain them, using the measurements of the radial velocities of the primary and
   secondary stars $K_1$ and $K_2$ in conjunction with the derived orbital period $P_{orb}$ and the spectral
   class of the secondary.

   Combining the values of $K_1$, $K_2$ and $P$ we find the masses of each component of the system
   \begin{equation}
    \label{M1sini}
      M_1 \sin^3 i = {P K_2 (K_1 + K_2)^2 \over 2 \pi G} = 0.34 \pm 0.01 M_{\odot}\,,
   \end{equation}
   \begin{equation}
    \label{M2sini}
      M_2 \sin^3 i = {P K_1 (K_1 + K_2)^2 \over 2 \pi G} = 0.14 \pm 0.01 M_{\odot}\,,
   \end{equation}
   \noindent and the projected binary separation
   \begin{equation}
      a \sin i = {P (K_1 + K_2) \over 2 \pi} = 1.37 \pm 0.02 R_{\odot}\,.
   \end{equation}

   The inclination angle $i$ of the system is an unknown parameter but for a CV it can be restricted by using some
   reasonable assumptions. The published photometry of BF~Eri is extensive enough to rule out any significant eclipse,
   which constrains the inclination $i$ to be less than about 70$^\circ$ to avoid an obvious partial eclipse of the disc.
   At $i$ = 70$^\circ$ the minimum WD mass is around 0.41 $M_\odot$. The Chandrasekhar mass limit determines the
   lowest limit for the inclination angle to be about 38$^\circ$. Thus, we have now the pure dynamic solution for the
   parameters of BF~Eri: $q \equiv M_2/M_1 = K_1/K_2 = 0.41 \pm 0.02$, $M_1$=0.41--1.44 $M_\odot$, $M_2$=0.17--0.59
   $M_\odot$, $i$=38$^\circ$--70$^\circ$, $a$=1.46--2.23 $R_{\odot}$.

\begin{figure}
\centering
\includegraphics[width=8cm]{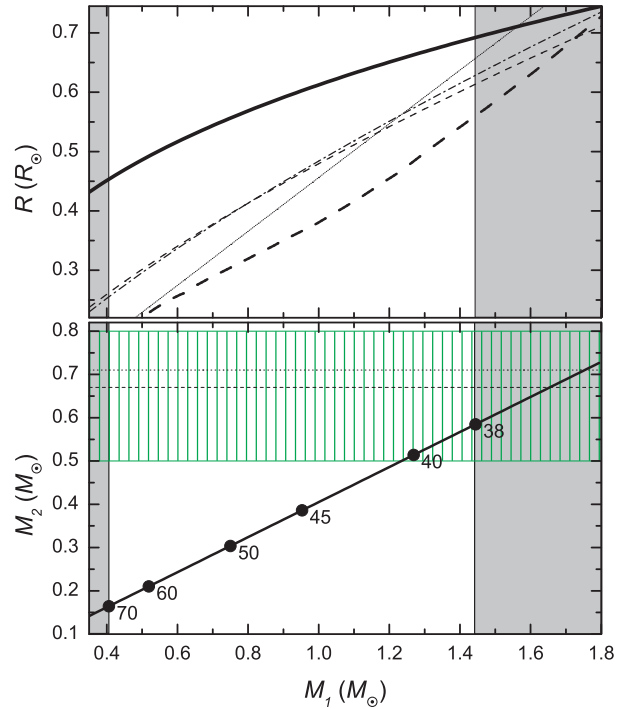}
\caption{
 Constraints on the system parameters of BF~Eri. In the bottom panel, we show the $M_1$ -- $M_2$ diagram.
 The diagonal solid line correspond to the derived mass ratio $q$=0.41. The dots along this line correspond to different
 mass solutions for specific values of the inclination angle, labelled nearby to those dots. The two vertical areas
 (shown in grey) mark the regions of impossible values, due to lack of eclipses (on the left) and the Chandrasekhar mass
 limit (on the right). The two horizontal lines show the predicted mass of the secondary from the $M_2$--$P_{orb}$
 relations by \citet{Warner} and \citet{Patterson05} (the dashed and dotted lines, respectively). The green vertical line
 pattern circumscribes the range of masses of the evolved main sequence star of the spectral type K3 under some
 \textit{extreme} assumptions as discussed by \citet{Kolb01}.
 In the upper panel we show, by the thick solid line, the Roche lobe size of the secondary $R$ as
 a function of the mass of the WD $M_1$, for the mass ratio $q$=0.41 and $P_{orb}$=0.270881 days.
 %The separation $a$ is calculated using Kepler's third law and shown in
 %accordance to the mass of the WD $M_1$ and the mass ratio $q$=0.41 (as shown by the diagonal solid line in the
 %bottom panel).
 The lower lines correspond to different theoretical and empirical Mass-Radius relations.
 The thick dashed line is the theoretical $M$-$R$ relation of \citet{BCAH98}. The thin dashed and dash-dotted lines
 are the empirical relations of \citet{Knigge} and \citet{Patterson05} respectively.
 The short-dotted line is the $M$-$R$ relation based on photometric, geometrical, and absolute elements for 112
 eclipsing detached binary systems with both components on the main sequence and with known photometric and
 spectroscopic orbital elements \citep{Gorda}.
  }
\label{fig:syspar}
\end{figure}

   At this point we would like to take note that from general theoretical considerations the donor star
   in a long-period CV should be more massive than in a short-period one. In practice, using any of the recently
   obtained empirical mass-period relations (see, for example,  \citealt{Warner, smith:dhillon, Patterson05})
   we obtain the mass for the secondary 0.67--0.72 $M_\odot$ (Fig.~\ref{fig:syspar},
   bottom panel). This is even more than our upper bound for the donor mass, leading us to the conclusion that
   the WD in BF~Eri is very close to the Chandrasekhar limit.

   Also we can estimate the secondary's mass, reasonably assuming that the secondary star fills its Roche lobe.
   The relative size of the donor star is constrained by Roche geometry and can be estimated using
   Eggleton's formula \citep{Eggleton}
   \begin{equation}
      \frac{R_2}{a} = \frac{0.49 q^{2/3}}{0.6 q^{2/3} + \ln(1+q^{1/3})},
   \end{equation}
   which gives the volume-equivalent radius of the Roche lobe to better than 1 per cent.

   In Fig.~\ref{fig:syspar} (upper panel) we show the Roche lobe size of the secondary as a function of
   the binary separation, and also different theoretical and empirical Mass-Radius relations. One can see that
   in order to fill its Roche lobe, the secondary must be relatively massive (0.7--0.8 $M_\odot$) if it is on the
   main sequence. But the mass of the primary in this case must exceed the Chandrasekhar limit, contradicting
   observations. A very massive white dwarf might just be accommodated again. However
   the secondary of the corresponding mass of 0.59 $M_\odot$ must be significantly larger than a main-sequence star of
   the same mass. It is now well known that all CVs have secondaries slightly larger ($\sim15$ per cent) than the
   zero-age main-sequence (ZAMS) single-star \citep{Patterson05, Knigge} but the secondary in BF~Eri is even greater
   by another 15 per cent.

   Additional constrains on the mass of the secondary can be obtained using another independently obtained parameter --
   its spectral type. Actually, this method should be used with great caution as it is understood that there is
   a big range in mass for a given spectral type \citep{smith:dhillon}. However, using the latest generation of low-mass
   star models, \citet{Kolb01} explored theoretically to what extent the spectral type provides a reasonable estimate
   of the donor mass. They concluded that the spectral type should be a good indicator of the donor mass, and
   gave lower and upper limits to $M_2$ for a given spectral type (see Table~1 in their paper). For the secondary star
   in BF~Eri we have the range of masses of $\sim 0.5-0.8 M_\odot$. The upper limit here is for the mass of a ZAMS star
   with the same spectral type as the donor while the lower limit is for evolved main sequence stars under some
   \textit{extreme} assumptions as discussed by \citet{Kolb01}. Thus, this leads us to conclude that the secondary in
   BF~Eri is an \textit{evolved} star of the mass of $\sim 0.50-0.59 M_\odot$. This in turn explains why the size
   of the secondary is so big for its mass. As \citet{Baraffe-Kolb-2000} showed,
   the size of the donor star of any given mass is larger for more evolved sequences.

   Thus, from the above analysis based on strong assumptions, we can restrict the system parameters
   of BF~Eri in the following ranges: $q$=0.41, $M_1$=1.23--1.44 $M_\odot$, $M_2$=0.50--0.59 $M_\odot$,
   $i$=38$^\circ$--40.5$^\circ$, $a$=2.11--2.23 $R_{\odot}$.
   Finally, in order to esimate the ``best-fitting'' values of the component masses and
   their errors, we used a Monte Carlo approach similar to \citet{Horne93}.
   The Monte Carlo simulation takes $10^6$ sample values of $K_1$ and $K_2$, treating each as being normally distributed
   about their measured values with standard deviations equal to the errors on the measurements. The values of the
   inclination $i$ are sampled at random from an uniform distribution over the interval [0, $i_{max}$], where
   $i_{max}=70^{\circ}$.
   Using the relations (\ref{M1sini}) and (\ref{M2sini}), we then
   calculate the masses of the components. If $M_1$ exceeds 1.44 $M_\odot$, or $M_2$ is less than 0.50 $M_\odot$, the
   outcome is rejected. The ensuing distribution for accepted outcomes is used to calculate the component masses.
   The values of all the system parameters deduced from the Monte Carlo simulation are listed in Table~\ref{Tab:Syspar}.
   Each value corresponds to the peak and standard deviation of the relevant distribution.

%************************  SysPar  ************************************
%************************  Chi2  ************************************

\begin{table}
\caption[] {Orbital and system parameters for BF Eridani.}
\begin{tabular}{lcc}
\hline
\hline\noalign{\smallskip}
Parameter                           & Measured                  & Monte Carlo  \\
                                    &   Value                   &     Value    \\
\noalign{\smallskip}
\hline\noalign{\smallskip}

$P_{orb}$ (d)                       & 0.270881$\pm3\times10^{-6}$\\
%T$_{0}$(set \textit{2005})          &  2453675.2670$\pm$0.???    \\
$T_{0}$ (+2450000)                  &  4061.8112$\pm$0.0003       \\
$K_1$ (\kms)                        &  74$\pm$3                 &\\
$K_2$ (\kms)                        &  182.5$\pm$0.9            &\\
$\gamma$ (\kms)                     &  -93.6$\pm$0.4            &\\
Spectral Type                       &                           &\\
       of secondary                 &  K3$\pm$0.5               &\\
$d$ (pc)                            &  700$\pm$200              &\\
$M_{1}/M_{\sun}$                    &  1.23--1.44               &  1.28$\pm$0.05    \\
$M_{2}/M_{\sun}$                    &  0.50--0.59               &  0.52$\pm$0.01    \\
$q=M_{2}/M_{1}$                     &  0.41$\pm$0.02            &  0.41$\pm$0.01    \\
$i$                                 &  38$^\circ$--40.5$^\circ$ &  40$^\circ\pm$1   \\
$a/R_{\sun}$                        &  2.11--2.23               &  2.14$\pm$0.02    \\
\noalign{\smallskip}
\hline

\end{tabular}
\label{Tab:Syspar}
\end{table}

\section{Discussion}
\label{DiscSec}

\subsection{System parameters}
\label{SysParDiscSect}

  We would like to make some final remarks on the system parameters of BF~Eri, derived in Section~\ref{SysParSec}.
  These parameters were determined using a traditional spectroscopic solution based on the derived radial velocity
  semi-amplitudes. In order to calculate the masses of the binary we have assumed that the semi-amplitudes of the
  measured radial velocities represent the true orbital motion of the stars. In CVs, however, this may not be an
  accurate assumption, as the emission lines arising from the disc may suffer several asymmetric distortions. Likewise,
  absorption line profiles may be subjected to irradiation or hot-spot contaminations in the surface of the secondary,
  distorting therefore the true value of $K_2$.

  Taking these potential sources of errors into consideration we, however, believe we could avoid them. Indeed,
  though those problems were severe in the early determinations of the radial velocities, nevertheless modern
  high resolution spectroscopy has greatly improved the methods in detecting asymmetric distortions to provide more
  reliable radial velocity values \citep{Warner}. In our study of BF~Eri, the value of $K_2$ is beyond question.
  Scattering of its values derived with the use of the different data, is very small. Doppler tomography does not
  show any emission from the secondary star that can distort the absorption line profiles from the latter, neither in
  quiescence nor in the outburst. Furthermore, a non-uniform absorption distribution across the surface of the secondary
  would result in a non-sinusoidal radial velocity curve that is not observed in BF~Eri (Fig.~\ref{FigRadVel}).
  All the obtained values of $K_1$ are also consistent with each other. Though the detected asymmetric emission
  structure of BF~Eri (particularly, the spot in the lower-left of the Doppler maps) may potentially influence
  the velocity determination, we believe we could avoid this as that strong emission spot is situated
  well inside of the chosen Gaussian separation. The correct phasing of the radial velocity curve also strengthens
  our confidence.

  Alternatively, one may also calculate the mass ratio of BF~Eri using the observed rotational velocity of the secondary
  star, independently of $K_1$. Unfortunately, we were unable to derive a consistent value of the rotational
  velocity, getting instead a broad range of \vsini\ values. This gives us the $q$ values in the range of
  0.33--0.62 that is consistent with the former solution, but also very uncertain, so we decided to reject it.

  Thus, we have shown that BF~Eri contains a massive WD and an evolved secondary.
  This places this system among a small group of CVs with a high-mass WD
  (such as RU~Peg and U~Sco). Such systems are specially interesting since they are considered as possible Type
  Ia supernova progenitors (\citealt{SNreview}, and references therein). The identity of the progenitor has become
  important for cosmology, since Type Ia supernovae are used as standard candles and their peak luminosity depends
  on the nature of the progenitor. The exploding star is now thought to be a carbon--oxygen WD
  that accretes mass from a binary companion until it approaches the Chandrasekhar limit, ignites carbon under
  electron-degenerate conditions, undergoes a thermonuclear instability, and disrupts completely.
  Further studies of the systems with massive WDs similar to BF~Eri are needed to understand if they evolve to SNe Ia.

\subsection{$\gamma$-velocity, proper motion and distance}

  A decade ago \citet{GamVel1} drew attention to the distribution of
  space velocities, as a means of probing the age profile of the CV population. \citet{KolbStehle} derived the
  theoretically expected distribution of $\gamma$-velocities and the dispersion of $\gamma$ as a function of orbital
  period. In particular, they showed that the CVs having periods longer than the upper limit of the period gap
  ($P_{orb}\geq$3 h) should be younger ($\leq$1.5 Gyr), and therefore have a smaller line-of-sight velocity dispersion
  according to the empirical age-velocity dispersion relation (predicted value $\sim$15 \kms). Conversely, those CVs with
  orbital periods shorter than the lower limit of the period gap, should be older ($\geq$ 3--4 Gyr) and show a larger
  velocity dispersion (predicted as $\sim$30 \kms). Paying a special attention to determine accurate absolute systemic
  velocities, \citet{North2002} found its values for four long-period DNs. They obtained an average
  $\gamma$-velocity of $\sim$5 \kms\ with a dispersion of $\sim$8 \kms\, indicating that the above postulate is
  indeed true.

  In our study of BF~Eri, we have mostly followed the method of \citet{North2002} and found the \textit{absolute}
  $\gamma$-velocity, after correction for the solar motion, to be $-94$ \kms, referred to the dynamical local
  standard of rest (LSR). This value is very far away from any expectations of the theory (as set out
  by \citealt{KolbStehle}). In principle, such a kick velocity could be caused by
  an event in which a significant fraction of the total mass of the system is violently, asymmetrically ejected.
  In the case of CVs, the progenitors indeed eject most of their mass during the common-envelope period, but it is
  supposed that this process does not alter the space velocity of the system. \citet{KolbStehle} note, however,
  that this velocity may be affected by the cumulative effect of repeated nova eruptions with
  asymmetric envelope ejection.

  It is worthy to mention that BF~Eri shows not only the high $\gamma$-velocity but also has a substantial
  proper motion: $(\mu_x, \mu_y) = (+34, -97)$ \masyr \citep{Klemola2000,Hanson2004}.
  In order to convert this value to the space velocity we need to know the distance to the system, which we can
  evaluate here by oblique methods only.

  One of these estimations is based on the \textit{statistical} period-luminosity-colours relation of \citet{Ak}.
  We have taken the $JHK$ magnitudes of BF~Eri from the 2MASS (Two Micron All Sky Survey)
  observations \citep{2MASS2,2MASS1} and found the distance to the system, from the dependence of the absolute
  magnitude $M_J$ on the orbital period $P_{orb}$ and colours $(J - H)$ and $(H - K_s)$, to be
  $700\pm200$ pc.

  Alternatively, the distances can be estimated based on a \textit{semi-empirical} donor sequence for CVs
  of \citet{Knigge}. Unfortunately, Knigge's donor sequence is intended to describe the
  unique evolution track followed by \textit{unevolved} secondaries, and he limited it to $P_{orb} \la 6.2$ h.
  BF~Eri is a system with a longer period and likely an evolved secondary. However,
  we have slightly extrapolated Knigge's sequence and found from it and the $JHK$ magnitudes
  the lower limit of the distance $d$ to be $\sim500$ pc. Applying the \textit{empirical} offsets
  to the donor absolute magnitudes (see, however, comments by Knigge on this) we have obtained $d$ to be $\sim900$ pc.

  Finaly, the distances can also be derived from the absolute magnitude at outburst $M_V$(max) versus the orbital period
  relation of \citet{Harrison}, if one intends to consider the flares of BF~Eri as DN outbursts (see corresponding
  discussion in Section~\ref{OutBehSec}). Analysis of the AAVSO database of BF~Eri's observations reveals 8 events brighter
  than 13 mag with an averaged value of $12.5 \pm 0.2$ mag. This yields the distance to be $650\pm60$ pc.

  Fortunately, all these independent estimations of the distance are consistent with each other, leading us to
  a final value of $700\pm200$ pc. This corresponds to a transverse velocity $v_T=340\pm100$ \kms\ or a space
  velocity of $\sim$350 \kms\ with respect to the LSR, that is again unusually high.
  Based on kinematics, \citet{BF-Eri} surmised that BF~Eri may belong to the halo population.
  Our Galactic components of the space velocity of the system $(U, V, W) = (-220, -280, 4)$ \kms are slightly different
  from those obtained by \citeauthor{BF-Eri} but also support a system motion being essentially in the Galactic plane and 
  lagging far behind the rotation of the Galactic disc. In this connection the referee has pointed out that the high space 
  velocity may be simply a sign of great age and not a result of any asymmetric mass ejections, that would be consistent 
  with the secondary star having started to evolve prior to the time the system came into contact. Indeed, though 
  \citet{KolbStehle} do mention asymmetric nova explosions as a possible mechanism for increasing the dispersion in the
  $\gamma$-velocity distribution, but they give no numbers, and it is not yet clear whether it is really plausible that the 
  effect could produce a $\gamma$-velocity as large as the one we found.

  However, if such a high space velocity of BF~Eri was indeed caused by repeated nova eruptions then it might be worth searching
  for its past (recurrent) nova outbursts in the astronomical archive. Although the system is known only since 1940 and
  during the following 50 years there were no observations, some collections of photographic plates cover over
  a century of data. It is straightforward to look at plates taken over the years to see if past eruptions
  have been recorded.

%************************  AAVSO1  ************************************
\begin{figure*}
\centering
\includegraphics[height=6.0cm]{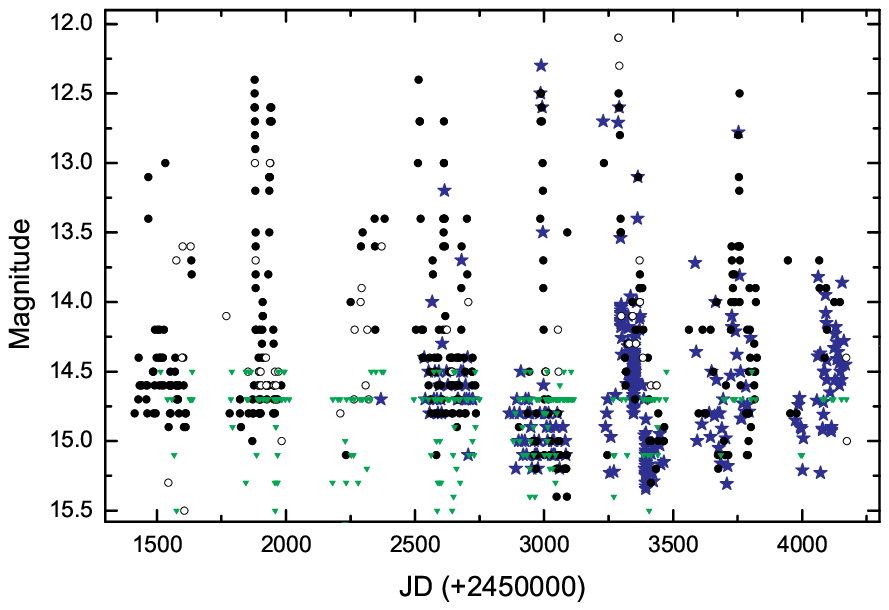}
\includegraphics[height=6.0cm]{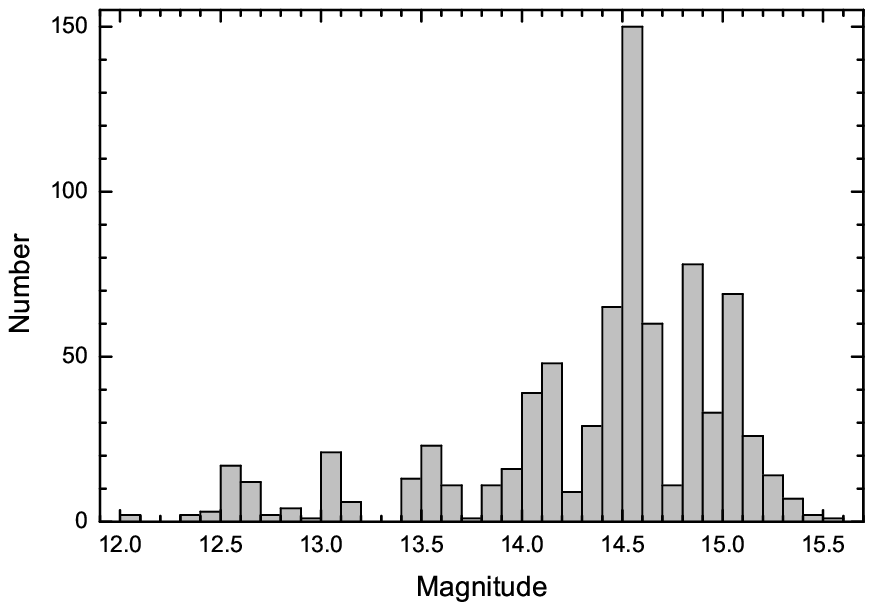}
\caption{\textbf{Left:} The AAVSO light curve of BF~Eri. Blue stars represent the data labelled in the
AAVSO database as V-magnitude, black filled dots - as Visual-magnitude, open dots - as inaccurate
Visual-magnitude, green triangles - as the ``Fainter Than'' Visual-data.
\textbf{Right:} Histogram of the reliable AAVSO data for BF~Eri.}
\label{fig:aavso1}
\end{figure*}
%************************  Chi2  ************************************

\subsection{The outburst behaviour and the nature of BF~Eri}
\label{OutBehSec}

  The current classification of BF~Eri as a DN is based on observations
  by Watanabe and Kato \citep{Watanabe, Kato99, KatoUemura} who detected ``outbursts''
  in the light curve of the system. We note that all those outbursts were of
  low amplitude (\textit{1--1.5 mag}) forcing Kato to classify BF~Eri as ``a low-amplitude dwarf nova''.
  However from the formal side, DNs are an subset of CVs, which undergo recurrent outbursts of
  \textit{2--8 mag}. The observed ``outbursts'', including the one observed during the \textit{2006-Nov} set,
  are of much lower amplitude, which leads us to the question: are those events the real DN outbursts rather
  than some kind of flares?

  It is now widely accepted that the reason for a DN outburst is a thermal instability in the accretion disc,
  which switches the disc from a low-viscosity to a high-viscosity regime \citep{Smak, Osaki, Lasota}. From the
  observational side this is (usually) accompanied by substantial spectral changes occurring continuously during all
  stages of the outburst. More than half of DNs exhibit the transition from emission-line spectrum at
  quiescence to absorption-line spectrum at maximum. For the rest if any emission can be seen it is often in the
  form of emission lines buried in absorption troughs \citep{Warner, MRM}. Quite often the spectral line profiles undergo
  dramatic changes as new emission sources arise in the system such as the inner hemisphere of the secondary star. This
  emission is caused by increased irradiation from the accretion regions during the outburst and can be seen in hydrogen
  and neutral helium lines through Doppler tomography \citep{Neustroev}. Many systems show also strengthening of
  the \HeII\ and \CIII/\NIII\ line emissions during an outburst.

  None of these outburst signs was observed in BF~Eri. Thus the observed flares can hardly be associated with a DN
  outburst. This in turn raises a question on the classification of the system. Actually, BF~Eri does not also show
  many other usual observational signs of DNs. For example, in spite of the high spectral resolution used, all
  the emission lines show not the double-peaked profiles but single-peaked ones. From the above-estimated system
  parameters we can expect the double peak separation to be more than 500 \kms, that is more than enough to be resolved.
  However, though the profiles are highly variable, they do not show any signs of the double-peaked accretion disc
  structure. In consequence, a ring-like emission distribution, characteristic of a Keplerian accretion disc about
  the WD, is also absent from the Doppler maps. On the other hand, the bulk of the Balmer emission is located to
  the lower left quadrant of the tomograms. Both the emission line profiles and the appearance of the Doppler maps
  resemble those in nova-like CVs which in turn resemble those typical for old novae at minimum light.

  By definition, NLs should not display any DN outbursts. Their almost steady brightness is thought to be due
  to their mass transfer rate $\dot{M}$ exceeding the upper stability limit $\dot{M}_{crit}$. These stars are thus
  thermally and tidally stable. However, \citet{Honeycutt98} reported that a significant fraction of NLs displays
  so-called ``stunted'' outbursts. \citet{Honeycutt01} found many similarities between them and outbursts in ordinary
  DNs but the former are of much smaller amplitude (0.4--1 mag). We suppose thus that BF~Eri might be a NL
  system exhibiting ``stunted'' outbursts.

  The AAVSO database has almost fourteen hundred BF~Eri observations obtained largely after 1999 \citep{AAVSO}.
  We have analyzed the light curve compiled from the `Visual magnitude' and `V magnitude' data and found that
  the outburst activity of BF~Eri is quite complex (Fig.~\ref{fig:aavso1}, left).
  Histogram of the AAVSO data reveals at least three
  types of flashes (Fig.~\ref{fig:aavso1}, right). Besides the above-mentioned low-amplitude flashes with the peak
  magnitude $\sim$13.0 mag and $\sim$13.5 mag, there were also 8 events with the amplitude more than 1.6 mag but always
  less than 2.3 mag, with $V_{max}\approx12$. Formally, the brightest flashes lie on the bottom edge of the accepted
  range of the outburst amplitudes in ordinary DNs and might be such outbursts. The nature of the
  low-amplitude flashes is still unclear; further investigation is needed.

\begin{figure}
\centering
\includegraphics{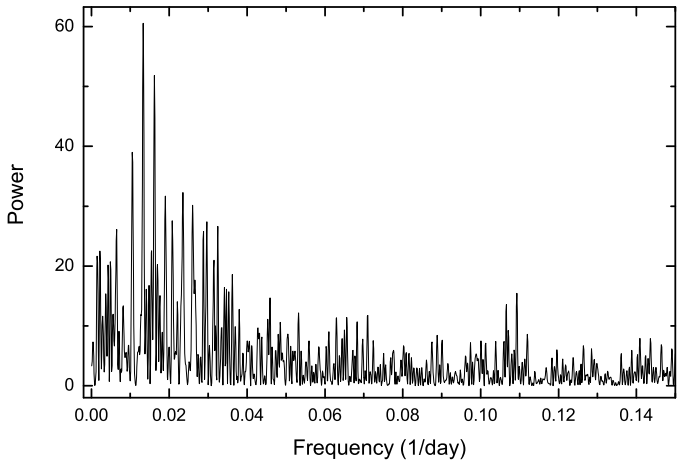}
\caption{The Lomb-Scargle power spectrum of BF~Eri, derived from the AAVSO data.}
\label{fig:aavso2}
\end{figure}

  Finally, we have obtained the periodogram, using the AAVSO data (Fig.~\ref{fig:aavso2}).
  It shows significant excess power in the interval 60--90 d, reflecting a reliable stability of the flashes.

%************************  AAVSO1  ************************************
%************************  Chi2  ************************************

\section{Summary}
\label{SumSec}

  We have presented spectroscopic observations of the cataclysmic variable BF~Eri during its low and bright states.
  The most important results of this study can be summarized as follows:
  \begin{enumerate}
    \item The orbital period of BF~Eri is $P_{orb}=0.270881(3)$ days.
    \item We have shown that BF~Eri contains a massive WD ($M_1\geq1.23 M_\odot$). This allows us to
       consider the system as a SN Ia progenitor.
    \item The spectral type of the secondary star is K3$\pm$0.5.
    \item We have shown that the secondary in BF~Eri is an evolved star of the mass of 0.50--0.59 $M_\odot$, whose
       size is about 30 per cent larger than a ZAMS star of the same mass.
    \item BF~Eri shows the high $\gamma$-velocity ($\gamma=-94$ \kms) and substantial proper motion.
       With our estimation of the distance to the system ($d\approx700\pm200$ pc), this corresponds to a space
       velocity of $\sim$350 \kms\ with respect to the LSR.
       The cumulative effect of repeated nova eruptions with asymmetric envelope ejection might explain the high space
       velocity of the system.
    \item We have analyzed the outburst behaviour of BF~Eri and question the current classification of
       the system as a dwarf nova. We propose that BF~Eri might be an old nova exhibiting ``stunted'' outbursts.
  \end{enumerate}

\section*{Acknowledgments}

    VN acknowledges support of IRCSET under their basic research programme and
    the support of the HEA funded CosmoGrid project.
    We would like to thank Tom Marsh for the use of his {\sc molly} software.
    This publication makes use of data products from the Two Micron All Sky Survey, which is a joint project of
    the University of Massachusetts and the Infrared Processing and Analysis Center/California Institute of Technology,
    funded by the National Aeronautics and Space Administration and the National Science Foundation.
    We acknowledge with thanks the variable star observations from the AAVSO
    International Database contributed by observers worldwide and used in this
    research.
    The authors would like to thank Gregg Hallinan for improving the language of the manuscript,
    and Gustavo Melgoza and Salvador Monrroy for their assistance during the observations.
    We are also grateful to the referee, R.C. Smith, for a very helpful report.
% The authors would like to thank Jean-Pierre Lasota and the referee for their useful comments,

\bsp
\label{lastpage}

\begin{thebibliography}{99}

  \bibitem[\protect\citeauthoryear{Ak et al.}{2007}]{Ak} Ak T., Bilir S., Ak S., Retter A., 2007, NewA, 12, 446
  \bibitem[\protect\citeauthoryear{Baraffe et al.}{1998}]{BCAH98} Baraffe I., Chabrier G., Allard F., Hauschildt P.~H.,
     1998, A\&A, 337, 403
  \bibitem[\protect\citeauthoryear{Baraffe \& Kolb}{2000}]{Baraffe-Kolb-2000} Baraffe I., Kolb U., 2000, MNRAS, 318, 354
%  \bibitem[\protect\citeauthoryear{Campbell \& Papaloizou}{1983}]{CampbellPapaloizou} Campbell C.~G., Papaloizou J.,
%     1983, MNRAS, 204, 433
  \bibitem[\protect\citeauthoryear{Chisholm et al.}{1999}]{Chisholm} Chisholm J.~R., Harnden F.~R., Jr., Schachter J.~F.,
     Micela G., Sciortino S., Favata F., 1999, AJ, 117, 1845
%  \bibitem[\protect\citeauthoryear{Czerny \& King}{1986}]{CzernyKing} Czerny M., King A.~R., 1986, MNRAS, 221, 55P
  \bibitem[\protect\citeauthoryear{Echevarr{\'{\i}}a et al.}{2007}]{Echevarria} Echevarr{\'{\i}}a J., Michel R.,
     Costero R., Zharikov S., 2007, A\&A, 462, 1069
  \bibitem[\protect\citeauthoryear{Eggleton}{1983}]{Eggleton} Eggleton P.~P., 1983, ApJ, 268, 368
  \bibitem[\protect\citeauthoryear{Elvis et al.}{1992}]{Elvis} Elvis M., Plummer D.,
     Schachter J., Fabbiano G., 1992, ApJS, 80, 257
  \bibitem[\protect\citeauthoryear{Gorda \& Svechnikov}{1998}]{Gorda} Gorda S.~Y., Svechnikov M.~A., 1998, ARep, 42, 793
  \bibitem[\protect\citeauthoryear{Hanley \& Shapley}{1940}]{HanleyShapley}
     Hanley C.~M., Shapley H., 1940, BHarO, 913, 9
  \bibitem[\protect\citeauthoryear{Hanson et al.}{2004}]{Hanson2004} Hanson R.~B., Klemola A.~R., Jones B.~F., Monet D.~G.,
     2004, AJ, 128, 1430
  \bibitem[\protect\citeauthoryear{Harrison et al.}{2004}]{Harrison} Harrison T.~E., Johnson J.~J., McArthur B.~E.,
     Benedict G.~F., Szkody P., Howell S.~B., Gelino D.~M., 2004, AJ, 127, 460
  \bibitem[\protect\citeauthoryear{Hellier \& Robinson}{1994}]{Hellier-Robinson} Hellier C., Robinson E.~L., 1994,
     ApJ, 431, L107
  \bibitem[\protect\citeauthoryear{Henden}{2007}]{AAVSO} Henden, A.A., 2007, Observations from the AAVSO International
     Database, private communication.
  \bibitem[\protect\citeauthoryear{Hoard et al.}{2002}]{2MASS2}Hoard D.~W., Wachter S., Clark L.~L.,
     Bowers T.~P., 2002, ApJ, 565, 511
  \bibitem[\protect\citeauthoryear{Honeycutt}{2001}]{Honeycutt01} Honeycutt R.~K., 2001, PASP, 113, 473
  \bibitem[\protect\citeauthoryear{Honeycutt, Robertson, \& Turner}{1998}]{Honeycutt98} Honeycutt R.~K.,
     Robertson J.~W., Turner G.~W., 1998, AJ, 115, 2527
  \bibitem[\protect\citeauthoryear{Horne, Welsh, \& Wade}{1993}]{Horne93} Horne K., Welsh W.~F., Wade R.~A., 1993,
     ApJ, 410, 357
  \bibitem[\protect\citeauthoryear{Kato}{1999}]{Kato99} Kato T., 1999, IBVS, 4745, 1
  \bibitem[\protect\citeauthoryear{Kato \& Uemura}{2000}]{KatoUemura}
     Kato T., Uemura M., 2000, IBVS, 4882, 1
  \bibitem[\protect\citeauthoryear{Keenan \& McNeil}{1976}]{Keenan-McNeil} Keenan P.~C., McNeil R.~C., 1976,
     An atlas of spectra of the cooler stars: Types G,K,M,S, and C. Part 1: Introduction and tables
     (Columbus: Ohio State University Press)
  \bibitem[\protect\citeauthoryear{Klemola, Hanson, \& Jones}{2000}]{Klemola2000} Klemola A.~R., Hanson R.~B.,
     Jones B.~F., 2000, Lick Northern Proper Motion Program: NPM1 Catalog (J2000 Version) (CDS Strasbourg Data
     Center Catalog No. I/199A), http://www.ucolick.org/~npm/NPM1
  \bibitem[\protect\citeauthoryear{Knigge}{2006}]{Knigge} Knigge C., 2006, MNRAS, 373, 484
  \bibitem[\protect\citeauthoryear{Kolb, King, \& Baraffe}{2001}]{Kolb01} Kolb U., King A.~R., Baraffe I., 2001,
     MNRAS, 321, 544
  \bibitem[\protect\citeauthoryear{Kolb \& Stehle}{1996}]{KolbStehle} Kolb U., Stehle R., 1996, MNRAS, 282, 1454
  \bibitem[\protect\citeauthoryear{Lasota}{2001}]{Lasota} Lasota J.-P., 2001, NewAR, 45, 449
  \bibitem[\protect\citeauthoryear{Levine \& Chakrabarty}{1995}]{Echelle} Levine S.,
     Chakrabarty D., 1995, IA-UNAM Technical Report \#MU-94-04
  \bibitem[\protect\citeauthoryear{Lomb}{1976}]{Lomb} Lomb N.~R., 1976, Ap\&SS, 39, 447
  \bibitem[\protect\citeauthoryear{Marsh}{2001}]{marsh2001} Marsh T.~R., 2001,
     in Astrotomography, Indirect Imaging Methods in Observational Astronomy, ed. H.~M.~J.~Boffin,
     D.~Steeghs, and J.~Cuypers, Lect. Notes Phys., 573, 1
  \bibitem[\protect\citeauthoryear{Marsh \& Horne}{1988}]{Marsh-Horne} Marsh T.~R., Horne K., 1988, MNRAS, 235, 269
  \bibitem[\protect\citeauthoryear{Marsh, Robinson, \& Wood}{1994}]{Marsh1994} Marsh T.~R., Robinson E.~L.,
     Wood J.~H., 1994, MNRAS, 266, 137
  \bibitem[\protect\citeauthoryear{Morales-Rueda \& Marsh}{2002}]{MRM} Morales-Rueda L., Marsh T.~R., 2002, MNRAS, 332, 814
  \bibitem[\protect\citeauthoryear{Neustroev, Zharikov, \& Michel}{2006}]{Neustroev} Neustroev V.~V., Zharikov S.,
     Michel R., 2006, MNRAS, 369, 369
  \bibitem[\protect\citeauthoryear{North et al.}{2000}]{North2000}North R.~C., Marsh T.~R.,
     Moran C.~K.~J., Kolb U., Smith R.~C., Stehle R., 2000, MNRAS, 313, 383
  \bibitem[\protect\citeauthoryear{North et al.}{2002}]{North2002}North R.~C., Marsh T.~R., Kolb U.,
     Dhillon V.~S., Moran C.~K.~J., 2002, MNRAS, 337, 1215
  \bibitem[\protect\citeauthoryear{Oke}{1990}]{Oke}Oke J.~B. 1990, AJ, 99, 1621
  \bibitem[\protect\citeauthoryear{Osaki}{1996}]{Osaki} Osaki Y., 1996, PASP, 108, 39
  \bibitem[\protect\citeauthoryear{Parthasarathy et al.}{2007}]{SNreview} Parthasarathy M., Branch D.,
     Jeffery D.~J., Baron E., 2007, NewAR, 51, 524
  \bibitem[\protect\citeauthoryear{Patterson et al.}{2005}]{Patterson05} Patterson J., et al., 2005, PASP, 117, 1204
  \bibitem[\protect\citeauthoryear{Roberts, Lehar, \& Dreher}{1987}]{CLEAN} Roberts D.~H., Lehar J., Dreher J.~W.,
     1987, AJ, 93, 968
  \bibitem[\protect\citeauthoryear{Scargle}{1982}]{Scargle} Scargle J.~D., 1982, ApJ, 263, 835
  \bibitem[\protect\citeauthoryear{Schachter et al.}{1996}]{Schachter}Schachter J.~F.,
     Remillard R., Saar S.~H., Favata F., Sciortino S., Barbera M., 1996, ApJ, 463, 747
  \bibitem[\protect\citeauthoryear{Schneider \& Young}{1980}]{sch:young} Schneider D.~P.,
     Young P., 1980, ApJ, 238, 946
  \bibitem[\protect\citeauthoryear{Shafter}{1983}]{Shafter} Shafter A.~W., 1983, ApJ, 267, 222
  \bibitem[\protect\citeauthoryear{Shafter \& Szkody}{1984}]{Shafter3} Shafter A.~W., Szkody P., 1984, ApJ 276, 305
  \bibitem[\protect\citeauthoryear{Shafter, Szkody, \& Thorstensen}{1986}]{Shafter2} Shafter A.~W.,
     Szkody P., Thorstensen J.~R., 1986, ApJ, 308, 765
  \bibitem[\protect\citeauthoryear{Sheets et al.}{2007}]{BF-Eri} Sheets H.~A., Thorstensen J.~R., Peters C.~J.,
     Kapusta A.~B., Taylor C.~J., 2007, PASP, 119, 494
  \bibitem[\protect\citeauthoryear{Skrutskie et al.}{2006}]{2MASS1} Skrutskie M.~F., et al., 2006, AJ,
     131, 1163
  \bibitem[\protect\citeauthoryear{Smak}{1984}]{Smak} Smak J., 1984, PASP, 96, 5
  \bibitem[\protect\citeauthoryear{Smith \& Dhillon}{1998}]{smith:dhillon} Smith D.~A., Dhillon V.~S., 1998,
     MNRAS, 301, 767
  \bibitem[\protect\citeauthoryear{Smith, Dhillon, \& Marsh}{1998}]{Smith1998} Smith D.~A.,
     Dhillon V.~S., Marsh T.~R., 1998, MNRAS, 296, 465
  \bibitem[\protect\citeauthoryear{Spruit}{1998}]{Spruit} Spruit H.~C., 1998, preprint (astro-ph/9806141)
  \bibitem[\protect\citeauthoryear{van Paradijs, Augusteijn, \& Stehle}{1996}]{GamVel1} van Paradijs J., Augusteijn T.,
     Stehle R., 1996, A\&A, 312, 93
  \bibitem[\protect\citeauthoryear{Warner}{1995}]{Warner} Warner, B., 1995,
     Cataclysmic Variable Stars (Cambridge Astrophysics Ser. 28; Cambridge: Cambridge Univ. Press)
  \bibitem[\protect\citeauthoryear{Watanabe}{1999}]{Watanabe} Watanabe, T., 1999, VSOLJ Variable Star Bulletin, 34


\end{thebibliography}
\end{document}